%% file: Reddy-Sysala2020.tex
\documentclass[a4paper, 12pt]{article}

\pdfoutput=1

\usepackage{color}
\usepackage{graphicx}
\usepackage{amsmath}
\usepackage{amsfonts}
\usepackage{amssymb}
\usepackage{amsthm}
\usepackage{epstopdf}
\usepackage{algorithmic}
\usepackage{setspace}
\usepackage[cm]{fullpage}

\newcommand{\mbf}[1]{\mbox{\boldmath$#1$}}

\newcommand{\sbpi}{\mbox{\boldmath{\scriptsize $\pi$}}}

\newcommand{\sbPi}{\mbox{\boldmath{\scriptsize $\Pi$}}}
\newcommand{\sbq}{\mbox{\boldmath{\scriptsize $q$}}}
\newcommand{\sbsigma}{\mbox{\boldmath{\scriptsize $\sigma$}}}
\newcommand{\sbx}{\mbox{\boldmath{\scriptsize $x$}}}

\numberwithin{equation}{section}

\input{defs.tex}

\begin{document}

\title{Bounds on the elastic threshold for problems of dissipative strain-gradient plasticity}
\author{B. D. Reddy\footnote{Department of Mathematics and Applied Mathematics, University of Cape Town, 7701 Rondebosch, South Africa. Email: daya.reddy@uct.ac.za. BDR acknowledges support for his work from the National Research Foundation, through the South African Chair in Computational Mechanics, SARChI Grant 47584.}\ \  and S. Sysala\footnote{Institute of Geonics of the Czech Academy of Sciences, Department of Applied Mathematics and Computer Science \& Department IT4Innovations, 708 00 Ostrava-Poruba, Czech Republic. E-mail: stanislav.sysala@ugn.cas.cz. SS acknowledges support for his work from the Czech Science Foundation (GA\v{C}R) through project No. 19-11441S.}}

\maketitle

{\bf Abstract}

This work is concerned with the purely dissipative version of a well-established model of rate-independent strain-gradient plasticity. In the conventional theory of plasticity the approach to determining plastic flow is local, and based on the stress distribution in the body. For the dissipative problem of strain-gradient plasticity such an approach is not valid as the yield function depends on microstresses that are not known in the elastic region. Instead, yield and plastic flow must be considered at the global level. This work addresses the problem of determining the elastic threshold by formulating primal and dual versions of the global problem and, motivated by techniques used in limit analysis for perfect plasticity, establishing conditions for lower and upper bounds to the threshold. The general approach is applied to two examples: of a plate under plane stress, and subjected to a prescribed displacement; and of a bar subjected to torsion.

{\bf Keywords:} dissipative strain-gradient plasticity, elastic threshold, duality, lower and upper bounds, penalization method, finite element method

\section{Introduction}

There has been sustained interest over the last few decades in the development and assessment of models of elastoplastic behaviour that take account of size-dependent responses at the microscale. A central and popular direction has been the development of models in which gradients of the plastic strain enter the formulation, as a means of accounting for the relationship between size-dependence and the development of geometrically necessary dislocations in the context of non-homogeneous deformations. Some representative examples of such strain-gradient theories include \cite{Fleck-Hutchinson2001,Gudmundson2001,Gurtin-Anand2004} (see also the references in these works).

The works \cite{Reddy_etal2008,Reddy2011a,Reddy2011b} have focused on analyses of well-posedness for particular formulations of strain-gradient plasticity. These analyses have given attention typically to what are referred to as energetic as well as dissipative models. Energetic formulations refer to models of strain-gradient plasticity in which the gradient terms arise from a recoverable free energy, and appear as a back-stress in the yield function. Dissipative formulations, on the other hand, are characterized by an extension of the conventional flow relation in which plastic strain rates as well as their gradients appear in the normality relation of associative plasticity. The two models lead respectively to size-dependent behaviour in the form of an increase in strain-hardening with length scale, and in the second case strengthening, that is, an increase in initial yield with increase in length scale. The works \cite{Chiricotto_etal2016a,Chiricotto_etal2016b} provide further examples of theoretical investigations of both kinds of behaviour.

Some features of rate-independent versions of dissipative problems have attracted particular attention. First, the generalization of the classical flow relation with a normality relation leads to a yield function expressed in terms of generalized microstresses which, because they are not known in the elastic region, cannot predict yield in the conventional way (see also \cite{Fleck-Willis2009}). On the other hand, it has been shown \cite{Reddy_etal2008,Reddy2011a} that the appropriate way to interpret yield and plastic flow lies through a global form of this relation, in which it then becomes possible to formulate the flow relation in terms of the Cauchy stress. This is most readily carried out for the kinematic form, using the dissipation function. In \cite{CEMRS17} the matter of determining a dual form of such a global flow relation is investigated further, where it is shown that a closed-form expression for such a relation, in terms of a global yield function, is elusive.

A second feature of dissipative problems relates to the response to non-proportional loading in the plastic range. This leads to an elastic gap: that is, an elastic response following the application of such loading. The phenomenon was first presented in \cite{Fleck-Hutchinson-Willis2014}, and has been further studied in various works, for example \cite{Bardella-Panteghini2015,Fleck-Willis2015,Martinez-Niordson-Bardella2016,McBride-Reddy-Steinmann2018,Panteghini-Bardella2016}. 

The objective of this work is to address, in the context of the dissipative problem of strain-gradient plasticity, the problem of determining the elastic threshold: that is, in relation to a scalar load factor, the stage at which incipient plastic behaviour takes place. The basis for the study is the dissipative version of the rate-independent theory presented and analyzed in \cite{CEMRS17,Reddy_etal2008,Reddy2011a}.

The approach taken in this work is to draw on the methods of limit analysis, a well-established area of investigation in conventional perfect plasticity. The link derives from the formulation, in limit analysis, of optimization problems that yield lower and upper bounds to collapse states \cite{T85,Ch96,HRS16,RSH18}. This correspondence is noted also in \cite[Section 7]{Fleck-Willis2009}, and is explored in detail in \cite{Polizzotto2010}, for a model in which size-dependence is through the gradient of a scalar function of plastic strain. In the current work it will be seen that the elastic threshold for problems of dissipative gradient plasticity may likewise be formulated as lower- and upper bound problems. 

The rest of this work is organized as follows. In Section \ref{sec_setting} the rate-independent problem of dissipative strain-gradient plasticity is presented, using the notions of a yield function and associative flow relation. The global form of the flow relation is derived in its kinematic form, that is, using the dissipation function or support function associated with the yield function. Section \ref{sec_yield_surface} makes precise the notion of plastically admissible stress fields in the context of strain-gradient plasticity. Inspired by classical limit load analysis for perfect plasticity, we obtain necessary and sufficient conditions for stress fields to be plastically admissible. This paves the way, in Section \ref{sec_threshold}, to derive expressions for lower and upper bounds to the elastic threshold, in terms of a parameter characterizing monotonic loading. Section \ref{sec_penalty}  introduces approximations in the form of penalizations of the highly nonlinear problems for the bounds. These theoretical results are applied to two example problems: in Section \ref{sec_example}, that of a square domain in plane stress and subjected to a prescribed displacement; and in Section \ref{sec_torsion}, the problem of torsion of a circular rod. For both problems approximations to the bounds are sought first by assuming simple forms for the minimizers or maximizers; and secondly, the problems are solved numerically, using finite element approximations in combination with penalized forms of the problems.

\section{The problem setting}
\label{sec_setting}
Vector- and 2nd-order tensor- or matrix-valued functions will be written in lower case boldface form. The scalar product of two tensors or matrices $\bsigma$ and $\btau$ will be denoted by $\bsigma:\btau$, and in component form, relative to an orthonormal basis, by $\sigma_{ij}\tau_{ij}$, the summation convention on repeated indices being invoked. Furthermore, we use upper case boldface letters to denote third-order tensors. For two such quantities $\bPi$ and $\bT$, the inner product is denoted by $\bPi\circ\bT$, or in component form $\Pi_{ijk}T_{ijk}$.

Consider a homogeneous elastic-plastic body that occupies a bounded domain $\Omega \in \BBR^3$ with boundary $\partial \Omega$. For a prescribed body force $\bb \in [L^2(\Omega)]^3$ the equilibrium equation reads
\begin{equation}
-\mathrm{div}\,\bsigma = \bb,
\label{balance1}
\end{equation}
where $\bsigma$ is the symmetric Cauchy stress tensor. The infinitesimal strain $\bepsilon$ is defined as a function of the displacement $\bu$ by
\begin{equation}
\bepsilon: = \bepsilon (\bu) = \mbox{$\frac{1}{2}$} (\nabla\bu+(\nabla\bu)^\top)\,.
\label{strain}
\end{equation}
The strain $\bepsilon$ is decomposed into elastic and plastic components $\bepsilon^e$ and $\bepsilon^p$, respectively, according to
\begin{equation}
\bepsilon = \bepsilon^e + \bepsilon^p\,.
\label{straindecomp}
\end{equation}
The material is assumed to be isotropic, with the elastic relation given by 
\begin{equation}
\bsigma = \BBC \bepsilon^e := \lambda (\mbox{tr}\,\bepsilon^e)\bI + 2\mu \bepsilon^e\,.
\label{Hooke}
\end{equation}
Here $\BBC$ is the fourth-order elasticity tensor, and $\lambda$ and $\mu$ are the two Lam\'e coefficients. 

This investigation is based on a model of strain-gradient plasticity \cite{Reddy2011a,Reddy_etal2008} in which the classical notions of a convex yield surface and normality flow relation are extended to the gradient case. This model takes as a point of departure the important work \cite{Gurtin-Anand2004}, which assumes viscoplastic behaviour without a yield surface. The theory makes use of second- and third-order microstresses $\bpi$ and $\bPi$ respectively. The quantity $\bpi$ is symmetric and deviatoric, and $\bPi$ is symmetric and deviatoric in its first two indices, in the sense that $\Pi_{ijk} = \Pi_{jik},\ \Pi_{ppk} = 0$. We will also require the deviatoric stress $\bsigma^D := \bsigma - \frac{1}{3}(\mbox{tr}\,\bsigma)\bI$.
In addition to the equation of macroscopic equilibrium the Cauchy stress and microstresses satisfy the microforce balance equation
\begin{equation}
\bsigma^D = \bpi - \mbox{div}\,\bPi\quad \mbox{or, in index form}\quad \sigma^D_{ij} = \pi_{ij} - \Pi_{ijk,k}\,.
\label{microforce_balance}
\end{equation}
The local form of the flow relation is posed in terms of a yield function $f_\ell (\bpi,\bPi)$, dependent on a length parameter $\ell$. The flow relation is then an extension of the classical normality relation and is given by
\begin{align}
& (\dot{\bepsilon}^p, \nabla \dot{\bepsilon}^p) = \gamma \left( \frac{\partial f_\ell}{\partial \bpi} , \frac{\partial f_\ell}{\partial \bPi}\right)\,,  \nonumber \\
& \gamma \geq 0,\quad f_\ell (\bpi,\bPi) \leq 0,\quad \gamma  f_\ell (\bpi,\bPi) = 0\,.
\label{flowlocal}
\end{align}
We will work with 
\begin{equation}
f_\ell(\mbf\pi,\mbf\Pi)=\sqrt{|\mbf\pi|^2+\ell^{-2}|\mbf\Pi|^2}-Y,
\label{yield}
\end{equation}
where $Y>0$ is the yield stress. It will be convenient also to introduce the local elastic region $E$, defined by
\begin{equation}
E  = \{(\bpi,\bPi)\ |\ f_\ell(\bpi,\bPi)\leq 0\}\,.
\label{Elocal}
\end{equation}

The flow relation may be recast with the microstresses as dependent variables by introducing the support function associated with $E$: in the present context this is referred to as the local dissipation function, denoted by $D_\ell$, and given by 
\begin{equation}
D_\ell (\bq,\nabla \bq) = \sup_{\substack{(\sbpi,\sbPi)\\[1pt] f_\ell(\sbpi,\sbPi)\leq0}}[\mbf\pi:\mbf q+\mbf\Pi\circ\nabla\mbf q]\,.
\label{Dlocal}
\end{equation}
For $f_{\ell}$ given by \eqref{yield}, the local dissipation function takes the form
\begin{equation}
D_\ell(\mbf q,\nabla\mbf q)=Y\sqrt{|\mbf q|^2+\ell^2|\nabla \mbf q|^2}\quad\mbox{in }\Omega.
\label{Dlocalexp}
\end{equation}
Then from a basic result in convex analysis, the flow relation \eqref{flowlocal} is equivalent to the inclusion
\begin{equation}
(\bpi,\bPi) \in \partial D_{\ell}(\dot{\bepsilon}^p,\nabla\bepsilon^p);
\label{primalflowlocal}
\end{equation}
here $\partial D_\ell$ is the subdifferential of $D_\ell$, so that \eqref{primalflowlocal} reads
\begin{equation}
D_\ell (\bq,\nabla \bq) \geq D_\ell (\dot{\bepsilon}^p,\nabla\bepsilon^p) + \bpi:(\bq - \dot\bepsilon^p) + \bPi\circ \nabla (\bq - \dot\bepsilon^p)\qquad \forall\ \ \bq\,.
\label{flowlocal2}
\end{equation} 

{\bf A note on alternative forms for $f_\ell$ and $D_\ell$.} A more general family of local dissipation functions may be defined by
\begin{equation}
D_{\ell,r} (\bq,\nabla \bq) : = Y[|\bq|^r + (\ell |\nabla\bq|)^r]^{1/r}
\label{D_ell^r}
\end{equation}
for real $r$ satisfying $1 \leq r < \infty$. The case $r=\infty$ corresponds to the function
\begin{equation}
D_{\ell,\infty} (\bq,\nabla \bq) : = Y\max \{ |\bq|,\ \ell |\nabla\bq|]\,.
\label{D_ell^infty}
\end{equation}
The yield function corresponding to $D_{\ell, r}$, $1<r\leq\infty$ reads
\begin{equation}
f_{\ell,r'} (\bpi,\bPi) : = \left[|\bpi|^{r'}+(\ell^{-1}|\bPi|)^{r'}\right]^{1/r'} - Y,\quad \frac{1}{r}+\frac{1}{r'}=1.
\label{f_ell^r}
\end{equation}
The case $r=1$ is also of practical significance \cite{Evans-Hutchinson2009}. The yield function $f_{\ell,\infty}$ corresponding to $D_{\ell,1}$ is shown in \cite{Reddy2011a} (see also \cite[Section 4.3.2]{Han-Reddy2013} to be given by
\begin{equation}
f_{\ell,\infty}(\bpi,\bPi) = \max\{ |\bpi|,\ \ell^{-1}|\bPi| \} - Y\,.
\label{f_ell_1}
\end{equation}


We will however focus in this work on the case $r = 2$, for which all significant features are likely to be present.

{\bf Initial and boundary conditions.}\ \ The initial conditions of the problem are
\begin{equation}
\bu = \bzero,\quad \bepsilon^p = \bzero\qquad \mbox{in}\ \Omega \times \{0\}\,.
\label{ics}
\end{equation}

For the boundary conditions associated with the equilibrium equation the boundary $\partial \Omega$ is partitioned into complementary parts $\partial\Omega_u$ and $\partial\Omega_t$ on which macroscopic Dirichlet and Neumann boundary conditions respectively are prescribed. These are
\begin{equation}
\mbf u=\bar{\mbf u}\;\;\mbox{on } \partial\Omega_u\times(0,T),\quad \mbf\sigma\mbf n=\bar{\mbf t}\;\;\mbox{on } \partial\Omega_t\times(0,T),
\label{bound_cond1}
\end{equation}
where $\bar{\mbf u}$ and $\bar{\mbf t}$ are respectively a prescribed displacement and traction, and $\mbf n$ denotes the outward unit normal to the boundary $\partial \Omega$.

Boundary conditions associated with the microforce balance equation are imposed on complementary parts
$\partial\Omega_H$ and $\partial\Omega_F$. The conditions on these two parts are referred to respectively as microhard and microfree boundary conditions, and are given by 
\begin{equation}
\mbf\varepsilon^p=\mbf 0\;\;\mbox{on } \partial\Omega_H\times(0,T),\quad \mbf\Pi\mbf n=\mbf 0\;\;\mbox{on } \partial\Omega_F\times(0,T).
\label{bound_cond2}
\end{equation}

The initial-boundary value problem for gradient plasticity is given by equations \eqref{balance1}, \eqref{straindecomp}, \eqref{strain}, \eqref{Hooke}, \eqref{microforce_balance}, \eqref{flowlocal} together with the initial and boundary conditions \eqref{ics}, \eqref{bound_cond1}, \eqref{bound_cond2}.

{\bf Remark.} \ \ The model presented here is a {\em dissipative} one, in the sense that recoverable energetic contributions to terms involving the plastic strain gradient are omitted. A treatment of the full problem may be found in \cite{Reddy2011a,Reddy_etal2008,Han-Reddy2013}. The structure of the dissipative model presents difficulties in its local form, and is approached as a global problem.

{\bf The global form of the flow relation.}\ \ It is clear from \eqref{flowlocal} or \eqref{flowlocal2} that, in contrast to the case of conventional plasticity, it is in fact not possible to use the local yield condition to determine pointwise whether the elastic limit has been reached. This is the case as $\bpi$  and $\bPi$ are not known a priori (cf. the case of a rigid-plastic body, for which it is not possible to determine the stresses in the rigid region, and hence to determine when yield occurs locally).

This dilemma may be resolved by approaching plastic flow as a global notion. For this purpose we construct a weak formulation of the microforce balance equation. We define the following space of tensor fields:
$$W=\{\mbf q\colon\Omega\rightarrow \mathbb R^{3\times 3}_{sym}\ |\;\; q_{ij}=q_{ji}, \; q_{ii}=0,\; q_{ij}\in L^2(\Omega),\; q_{ij,k}\in L^2(\Omega),\; \mbf q=\mbf 0 \;\mbox{on } \partial\Omega_H\}.$$
The $L^2$-space of all pairs $(\mbf\pi,\mbf\Pi)$ will be denoted by $X$.
We take the inner product of  \eqref{microforce_balance} with $\bq - \dot{\bepsilon}^p \in W$, integrate, integrate the term involving $\bPi$ by parts, and use the boundary condition \eqref{bound_cond2}$_2$, to obtain
\begin{equation}
\int_\Omega [\bpi:(\bq-\dot{\bepsilon}^p)+\bPi\circ\nabla (\bq - \dot{\bepsilon}^p)]\,dx=
\int_\Omega \bsigma^D:(\bq - \dot{\bepsilon}^p)\,dx\quad\forall\ \bq\in W.
\label{weak_balance}
\end{equation}
Next, integrate \eqref{flowlocal2} and use \eqref{weak_balance} to eliminate the terms involving $\bpi$ and $\bPi$ from the resulting global inequality. The result is the global flow relation
\begin{equation}
\int_\Omega D_\ell (\bq,\nabla \bq)\ dx \geq \int_\Omega D_\ell (\dot{\bepsilon}^p,\nabla\bepsilon^p)\ dx + \int_\Omega \bsigma:(\bq - \dot\bepsilon^p)\ dx \qquad \forall\ \bq \in W\,.
\label{globalflow}
\end{equation}

{\bf Remark.} The variational inequality \eqref{globalflow} together with the weak form of the equilibrium equation has a unique solution in appropriate Sobolev spaces provided that there is some hardening present \cite{Reddy2011a}. For example, kinematic hardening would result in the replacement of the second term on the righthand side of \eqref{globalflow} by $\bsigma - \kappa \bepsilon^p$, where $\kappa \in L^\infty(\Omega)$ is required to satisfy $\kappa (\bx) \geq \bar{\kappa} > 0$. We will omit the hardening term in what follows, as it will not be central to the focus in what follows, viz. a global form of the flow relation.

\section{Plastically admissible stresses}
\label{sec_yield_surface}

Within this section, we fix $t\in(0,T)$. From the classical problem \eqref{balance1}--\eqref{flowlocal}, \eqref{ics}--\eqref{bound_cond2} we say that the stress field $\mbf\sigma$ is {\it plastically admissible} if there exists a pair $(\mbf \pi,\mbf \Pi)$ satisfying
\begin{equation}
\left\{\begin{array}{c c}
\mbf \pi-\mathrm{div}\,\mbf \Pi=\mbf\sigma^D &\mbox{in }\Omega,\\
\mbf \Pi \mbf n=\mbf 0 &\mbox{on }\partial\Omega_F,\\
f_\ell(\mbf \pi,\mbf \Pi)\leq0\ &\mbox{in }\Omega.
\end{array}\right.
\label{E_glob3}
\end{equation}
We denote the set of all plastically admissible stresses by $\mathcal E_{glob}^s$. As indicated earlier, the conditions in \eqref{E_glob3} cannot be verified locally by a direct approach, and so an alternative approach has to be adopted to determine whether $\mbf\sigma\in\mathcal E_{glob}^s$.

Let $\mbf\Pi$ be such that (\ref{E_glob3})$_2$ holds and $\mathrm{div}\,\mbf \Pi$ exists. Inserting (\ref{E_glob3})$_1$ into (\ref{E_glob3})$_3$, we arrive at the inequality 
\begin{equation}
f_\ell(\mbf\sigma^D+\mathrm{div}\,\mbf \Pi,\mbf \Pi)\leq0\;\;\mbox{in }\Omega,
\label{suff_cond1}
\end{equation}
which is a {\it sufficient} condition to be $\mbf\sigma\in\mathcal E_{glob}^s$. In particular, if $\mbf \Pi=\mbf 0$ then (\ref{suff_cond1}) reduces to
\begin{equation}
f_\ell(\mbf\sigma^D,\mbf 0)\leq0\;\;\mbox{in }\Omega.
\label{suff_cond2}
\end{equation}
Thus local yield is a sufficient condition for $\bsigma$ to be plastically admissible.

To find {\it necessary} conditions for $\mbf\sigma\in\mathcal E_{glob}^s$, we develop an alternative approach to the definition of plastically admissible stresses using duality theory. The definition based on (\ref{E_glob3}) may be interpreted as the static approach while the dual definition will be kinematic in nature. The following derivation is inspired by the techniques of limit analysis in perfect plasticity \cite{T85, Ch96, HRS19}.

First, define the auxiliary function $\tilde\varphi_{glob}(\mbf\sigma):=$ {\it supremum over all $\lambda\geq0$ for which there exists $(\mbf \pi,\mbf \Pi)$ such that}
\begin{subequations}
\begin{align}
\mbf \pi-\mathrm{div}\,\mbf \Pi & = \lambda\mbf\sigma^D \quad \mbox{in }\Omega, \label{weakbal2}\\
\mbf \Pi \mbf n & = \mbf 0  \quad \quad \mbox{on }\partial\Omega_F, \label{microfree2}\\
f_\ell(\mbf \pi,\mbf \Pi) & \leq0  \quad \quad \mbox{in }\Omega. 
\label{phi3}
\end{align}
\end{subequations}
Let $\mathcal E_{glob}^k:=\left\{\mbf\sigma\ |\;\; \tilde\varphi_{glob}(\mbf\sigma)\geq 1\right\}$. It is readily seen that the inequality $\tilde\varphi_{glob}(\mbf\sigma)\geq 1$ is a necessary condition to be $\mbf\sigma\in\mathcal E_{glob}^s$ and that $\mathcal E_{glob}^s\subset \mathcal E_{glob}^k$. In addition, if $\tilde\varphi_{glob}(\mbf\sigma)> 1$ then (\ref{phi3}) holds for any $\lambda<\tilde\varphi_{glob}(\mbf\sigma)$, i.e., for $\lambda=1$ implying $\mbf\sigma\in\mathcal E_{glob}^s$. On the other hand, if $\tilde\varphi_{glob}(\mbf\sigma)=1$ then it is not clear whether $\mbf\sigma$ belongs to $\mathcal E_{glob}^s$ or not. This task is beyond the scope of this work.

From \eqref{weak_balance} the weak form of \eqref{weakbal2}--\eqref{microfree2} reads  
\begin{equation}
\int_\Omega [\mbf\pi:\mbf q+\mbf\Pi\circ\nabla\mbf q]\,dx=\lambda\int_\Omega \mbf\sigma^D:\mbf q\,dx\quad\forall \mbf q\in W.
\label{weak_balance_lambda}
\end{equation}
This condition holds if and only if
\begin{equation}
\lambda=\inf_{\substack{\sbq\in W \\[1pt] \int_\Omega \sbsigma^D:\sbq\,dx=1}}\int_\Omega [\mbf\pi:\mbf q+\mbf\Pi\circ\nabla\mbf q]\,dx.
\label{equivalence}
\end{equation}
Hence, we arrive at
\begin{equation}
\tilde\varphi_{glob}(\mbf\sigma)=\sup_{\substack{(\sbpi,\sbPi)\in X\\[1pt] f_\ell(\sbpi,\sbPi)\leq0\, \mathrm{in}\, \Omega}}\ \inf_{\substack{\sbq\in W \\[1pt] \int_\Omega \sbsigma^D:\sbq\,dx=1}}\int_\Omega [\mbf\pi:\mbf q+\mbf\Pi\circ\nabla\mbf q]\,dx.
\label{phi2}
\end{equation}

Next, we swop the order of inf and sup in (\ref{phi2}). In general, we have $\sup\, \inf\leq \inf\, \sup$. Sufficient conditions for $\sup\, \inf= \inf\, \sup$ are introduced e.g. in \cite[Chapter VI]{ET74}. For the present problem, from \cite[Proposition VI.2.3 and Remark VI.2.3]{ET74} it follows that the equality $\sup\, \inf= \inf\, \sup$ holds if the set $\{(\mbf\pi,\mbf\Pi)\in X\ |\; f_\ell(\mbf\pi,\mbf\Pi)\leq0\; \mbox{in } \Omega\}$ is bounded. This is the case for the function $f_\ell$ introduced in (\ref{yield}) and considered in this work. Therefore, we can write
\begin{equation}
\tilde\varphi_{glob}(\mbf\sigma)=\inf_{\substack{\sbq\in W \\[1pt] \int_\Omega \sbsigma^D:\sbq\,dx=1}}\ \sup_{\substack{(\sbpi,\sbPi)\in X\\[1pt] f_\ell(\sbpi,\sbPi)\leq0\, \mathrm{in}\, \Omega}}\ \int_\Omega [\mbf\pi:\mbf q+\mbf\Pi\circ\nabla\mbf q]\,dx.
\label{phi}
\end{equation}
Using \cite[Proposition IX.2.1]{ET74}, it is possible also to swop the supremum and the integral in (\ref{phi}):
\begin{equation}
\sup_{\substack{(\sbpi,\sbPi)\in X\\[1pt] f_\ell(\sbpi,\sbPi)\leq0\, \mathrm{in}\, \Omega}}\int_\Omega [\mbf\pi:\mbf q+\mbf\Pi\circ\nabla\mbf q]\,dx=\int_\Omega \sup_{\substack{(\sbpi,\sbPi)\in X\\[1pt] f_\ell(\sbpi,\sbPi)\leq0}}[\mbf\pi:\mbf q+\mbf\Pi\circ\nabla\mbf q]\,dx=\int_\Omega D_\ell(\mbf q,\nabla\mbf q)\,dx,
\label{j}
\end{equation}
where $D_\ell$ is the dissipation introduced in (\ref{Dlocal}) or (\ref{Dlocalexp}). Hence,
\begin{equation}
\tilde\varphi_{glob}(\mbf\sigma)=\inf_{\substack{\sbq\in W \\[1pt] \int_\Omega \sbsigma^D:\sbq\,dx=1}}\ \int_\Omega D_\ell(\mbf q,\nabla\mbf q)\,dx.
\label{phi1}
\end{equation}
Using (\ref{phi1}), the (kinematic) global yield surface in gradient plasticity is defined as follows:
\begin{equation}
\mathcal E_{glob}^k=\left\{\mbf\sigma\ |\;\; \tilde\varphi_{glob}(\mbf\sigma)\geq 1\right\}=\left\{\mbf\sigma\ |\;\; \int_\Omega \mbf\sigma^D:\mbf q\,dx\leq \int_\Omega D_\ell(\mbf q,\nabla\mbf q)\,dx\;\;\;\forall \mbf q\in W\right\}.
\label{E_glob}
\end{equation}
If $\mbf\sigma\in \mathcal E_{glob}^s$ then the inequality in (\ref{E_glob})
must be satisfied for any $\mbf q\in W$. That is, this inequality is a necessary condition for $\mbf\sigma$ to be plastically admissible. The relation \eqref{E_glob} has also been obtained in \cite[Secton 6.4.1]{McBride-Reddy-Steinmann2018}, using arguments based on polar conjugates of convex positively homogeneous functionals.

\section{The elastic threshold and its lower and upper bounds}
\label{sec_threshold}

In this section we assume that the data (that is, $\bar{\mbf u}$, $\bar{\mbf t}$ and $\mbf b$ in Section \ref{sec_setting}) depend linearly on the time variable $t$; that is,
$$\bar{\mbf u}=t\hat{\mbf u},\quad \bar{\mbf t}=t\hat{\mbf t},\quad \mbf b=t\hat{\mbf b},\quad t\in(0,T),$$
where $\hat{\mbf u}$, $\hat{\mbf t}$ and $\hat{\mbf b}$ are given functions defined in $\Omega$. We consider the evolution of the solution to the problem \eqref{balance1}--\eqref{flowlocal}, \eqref{ics}--\eqref{bound_cond2} over the time interval $(0,T)$. Clearly, for sufficiently small $t > 0$ this problem has an elastic solution of the form
\begin{equation}
\mbf u=t\hat{\mbf u}_e,\quad \mbf\varepsilon=t\hat{\mbf\varepsilon}_e,\quad \mbf\sigma=t\hat{\mbf\sigma}_e,\quad\mbf\varepsilon^p=\mbf0,\quad\gamma=0,\quad \mbf\pi=t\hat{\mbf\pi}_e,\quad \mbf\Pi=t\hat{\mbf\Pi}_e.
\label{elast_branch}
\end{equation}
The solution components $\hat{\mbf\pi}_e$ and $\hat{\mbf\Pi}_e$ need not be uniquely defined as we shall see in Section \ref{sec_example}.

The {\it elastic threshold} $t^*$ is the maximal value of $t$ for which the solution to \eqref{balance1}--\eqref{flowlocal}, \eqref{ics}--\eqref{bound_cond2} satisfies (\ref{elast_branch}). It can be defined through the plastically admissible stress fields: that is,
\begin{align}
t^* &:=\max\big\{t>0\ |\;\; \exists(\mbf\pi,\mbf\Pi)\in X:\;\; t\hat{\mbf\sigma}_e^D=\mbf \pi-\mathrm{div}\,\mbf\Pi,\nonumber\\ &\qquad f_\ell(\mbf\pi,\mbf\Pi)\leq0\;\;\mbox{in }\Omega,\;\;\mbf \Pi \mbf n=\mbf 0\;\;\mbox{on }\partial\Omega_F\big\}.
\label{t*1}
\end{align}
The results of Section \ref{sec_yield_surface} can be applied to find lower and upper bounds of $t^*$. With $f_\ell$ defined by (\ref{yield}), \eqref{t*1} can be written 
\begin{align}
t^*& =\max\big\{t>0\ |\;\; \exists(\mbf\pi,\mbf\Pi)\in X:\;\; \hat{\mbf\sigma}_e^D=\mbf \pi-\mathrm{div}\,\mbf\Pi,\nonumber \\
& \qquad  t\sqrt{|\mbf\pi|^2+\ell^{-2}|\mbf\Pi|^2}\leq Y\;\;\mbox{in }\Omega, \ \mbf\Pi\mbf n= \mbf 0\;\;\mbox{on }\partial\Omega_F\big\}\,.
\label{t*0}
\end{align}
The lower bound $t^*_L$ of $t^*$ can then be written in terms of $\mbf\Pi$ and $\hat{\mbf\sigma}^D_e$ by eliminating $\mbf\pi$ from \eqref{t*0}: this gives
\begin{equation}
t^*_L=\frac{Y}{\min\limits_{\sbPi}\Big\|\sqrt{|\hat{\mbf\sigma}^D_e+\mathrm{div}\,\mbf \Pi|^2+\ell^{-2}|\mbf\Pi|^2}\Big\|_{L^\infty(\Omega)}},
\label{lower_bound1}
\end{equation}
where $\mbf\Pi$ is a third-order tensor-valued function such that $\mbf \Pi \mbf n=\mbf 0\;\mbox{on }\partial\Omega_F$ and $\mathrm{div}\,\mbf \Pi$ exists in an appropriate sense. 
For a sufficiently sharp bound it is necessary 
to minimize the denominator in (\ref{lower_bound1}) with respect to  $\mbf \Pi$. 
On the other hand, setting $\mbf\Pi=\mbf0$, (\ref{lower_bound1}) reduces to 
\begin{equation}
t^*_{L,1}=Y/\big\||\hat{\mbf\sigma}^D_e|\big\|_{L^\infty(\Omega)}
\label{lower_bound2}
\end{equation}
(cf. eqn (5.15) in \cite{CEMRS17} for a similar result in the discrete context). We note that $t^*_{L,1}$ defines the elastic threshold in perfect (Hencky) plasticity.

The ``kinematic" definition of $t^*$ reads 
\begin{equation}
t^*=\tilde\varphi_{glob}(\hat{\mbf\sigma})=\inf_{\substack{\sbq\in W \\[1pt] \int_\Omega \hat{\sbsigma}^D_e:\sbq\,dx=1}}\ \int_\Omega D_\ell(\mbf q,\nabla\mbf q)\,dx=\inf_{\substack{\sbq\in W \\[1pt] \int_\Omega \hat{\sbsigma}^D_e:\sbq\,dx=1}}\ \int_\Omega Y\sqrt{|\mbf q|^2+\ell^2|\nabla \mbf q|^2}\,dx.
\label{t*2}
\end{equation}
It enables us to find the upper bounds of $t^*$ from
\begin{equation}
t^*_{U}=\frac{\displaystyle \int_\Omega Y\sqrt{|\mbf q|^2+\ell^2|\nabla \mbf q|^2}\,dx}{\displaystyle \int_\Omega \hat{\mbf \sigma}^D_e:\mbf q\,dx},
\label{upper_bound}
\end{equation}
for any $\mbf q\in W$ such that $\int_\Omega \hat{\mbf \sigma}^D_e:\mbf q\,dx>0$.

\subsection{The elastic threshold for a constant stress field}
\label{subsec_constant}

We consider here a subclass of problems for which the stress field $\hat{\mbf \sigma}_e$ is constant in $\Omega$. For this case the formulas  
(\ref{t*1}) and (\ref{t*2}) defining the elastic threshold $t^*$ can be simplified. To this end, we consider the following forms of $\mbf\pi$, $\mbf\Pi$ and $\mbf q$:
\begin{equation}
\mbf\pi=t\hat{\mbf \sigma}_e^D\tilde\pi,\quad  \mbf\Pi=t\hat{\mbf \sigma}_e^D\otimes\tilde{\mbf\Pi},\quad \mbf q=\hat{\mbf \sigma}_e^D\tilde{q}\,.
\label{piPiq}
\end{equation}
Here $\tilde\pi,\tilde q\in L^2(\Omega)$, $\tilde{\mbf\Pi}\in L^2(\Omega;\mathbb R^3)$, and $[\hat{\mbf \sigma}_e^D\otimes\tilde{\mbf\Pi}]_{ijk}=[\hat{\mbf \sigma}_e^D]_{ij}[\tilde{\mbf\Pi}]_{k}$. Inserting (\ref{piPiq}) into (\ref{t*1}) and (\ref{t*2}), we obtain
\begin{equation}
\frac{Y}{|\hat{\mbf \sigma}_e^D|}\tilde t^*_1\leq t^*\leq \frac{Y}{|\hat{\mbf \sigma}_e^D|}\tilde t^*_2,
\label{t*3}
\end{equation}
where
\begin{align}
\tilde t^*_1 & =\max\big\{t>0\ |\;\; \exists(\tilde\pi,\tilde{\mbf\Pi})\in \tilde X:\;\; 1=\tilde \pi-\mathrm{div}\,\tilde{\mbf\Pi},\nonumber\\
& \qquad t\sqrt{\tilde\pi^2+\ell^{-2}|\tilde{\mbf\Pi}|^2}\leq1\;\;\mbox{in }\Omega,\;\;\tilde{\mbf\Pi}\cdot\mbf n= 0\;\;\mbox{on }\partial\Omega_F\big\},
\label{t*L}
\end{align}
and 
\begin{equation}
\tilde t^*_2=\inf_{\substack{\tilde q\in \tilde W \\[1pt] \int_\Omega \tilde q\,dx=1}}\ \int_\Omega \sqrt{\tilde q^2+\ell^2|\nabla \tilde q|^2}\,dx,\quad \tilde W=\{\tilde q\in W^{1,2}(\Omega)\ |\;\; \tilde q=0\;\mbox{on } \partial\Omega_H\}.
\label{t*U}
\end{equation}
Here and henceforth $W^{m,p}(\Omega)$ denotes the Sobolev space of functions which together with their weak derivatives of order $\leq m$ are in $L^p(\Omega)$, for integer $p\geq 1$.
Analogously as in Section \ref{sec_yield_surface}, one can derive the duality between (\ref{t*L}) and (\ref{t*U}), implying $\tilde t^*_1=\tilde t^*_2=:\tilde t^*$. Hence,
\begin{equation}
t^*= \frac{Y}{|\hat{\mbf \sigma}_e^D|}\tilde t^*,
\label{t*4}
\end{equation}
where $\tilde t^*$ is defined either by (\ref{t*L}) or (\ref{t*U}). The lower and upper bounds (\ref{lower_bound1}) and (\ref{upper_bound}) of $t^*$ may be rewritten as follows:
\begin{equation}
t^*_L=\frac{Y}{|\hat{\mbf\sigma}^D_e|}\frac{1}{\Big\|\sqrt{(1+\mathrm{div}\,\tilde{\mbf \Pi})^2+\ell^{-2}|\tilde{\mbf\Pi}|^2}\Big\|_{L^\infty(\Omega)}},\quad \tilde{\mbf\Pi}\in \tilde X_\Pi,
\label{lower_bound3}
\end{equation}
and
\begin{equation}
t^*_{U}=\frac{Y}{|\hat{\mbf\sigma}^D_e|}\frac{\displaystyle \int_\Omega \sqrt{\tilde q^2+\ell^2|\nabla \tilde q|^2}\,dx}{\int_\Omega \tilde q\,dx},\quad \tilde q\in\tilde W,
\label{upper_bound3}
\end{equation}
respectively, where
$$\tilde X_\Pi=\{\tilde{\mbf\Pi}\in L^\infty(\Omega;\mathbb R^3)\;|\;\; \mathrm{div}\,\tilde{\mbf \Pi}\in L^\infty(\Omega;\mathbb R^3),\;\; \tilde{\mbf\Pi}\cdot\mbf n=0\;\mbox{on }\partial\Omega_F\}.$$

\section{Penalization methods for estimating the elastic threshold}
\label{sec_penalty}

The elastic threshold $t^*$ may be estimated according to the results presented in the previous section. Our aim is to find lower and upper bounds $t^*_L$ and $t^*_U$ which are sufficiently close to $t^*$. Let us recall that the bounds $t^*_L$ and $t^*_U$ are defined by appropriate functions $\mbf \Pi$ and $\mbf q$, respectively. In some special cases, it is sufficient to choose $\mbf \Pi$ and $\mbf q$ by analytical formulas. Nevertheless, the usage of numerical methods is more widely applicable.
	
To find numerical solutions, it is necessary to discretize the problems \eqref{lower_bound1} and \eqref{t*2}. If we choose finite-dimensional spaces $X_h$ and $W_h$ such that $X_h\subset X$ and $W_h\subset W$ then any admissible $\mbf \Pi_h\in X_h$ and $\mbf q_h\in W_h$ define lower and upper bounds of $t^*$, respectively.

Next, it is difficult and/or slow to solve the problems \eqref{lower_bound1} and \eqref{t*2} or their discrete counterparts directly because they contain non-differentiable functionals. Therefore, it is convenient to transform or modify these problems. To this end, we make use of penalization methods. We describe the methods in the context of the simplified problem from Section \ref{subsec_constant}, corresponding to a constant stress field. 

\subsection{Penalization method for the lower-bound problem}
\label{subsec_penalty1}

Returning to equation \eqref{lower_bound3} in Section \ref{subsec_constant}, consider the following lower bound problem:
\begin{equation}
s^*:=\inf_{\tilde\sbPi\in\tilde X_\Pi}\mathcal I(\tilde{\mbf\Pi}),\qquad \mathcal I(\tilde{\mbf\Pi})=\left\|(1+\mathrm{div}\,\tilde{\mbf\Pi})^2+\ell^{-2}|\tilde{\mbf\Pi}|^2\right\|_{L^\infty(\Omega)},
\label{q*}
\end{equation}
where 
$$\tilde X_\Pi=\{\tilde{\mbf\Pi}\in L^\infty(\Omega;\mathbb R^3)\;|\;\; \mathrm{div}\,\tilde{\mbf \Pi}\in L^\infty(\Omega;\mathbb R^3),\;\; \tilde{\mbf\Pi}\cdot\mbf n=0\;\mbox{on }\partial\Omega_F\}.$$
Under the assumptions in Section \ref{subsec_constant}, we have:
$$t^*=\frac{Y}{|\hat{\mbf\sigma}_e^D|}\frac{1}{\sqrt{s^*}},\quad t^*_L=\frac{Y}{|\hat{\mbf\sigma}_e^D|}\frac{1}{\sqrt{\mathcal I(\tilde{\mbf\Pi})}}\leq t^*\quad\forall \tilde\Pi\in \tilde X_\Pi.$$

For penalization of the problem \eqref{q*}, we replace the $L^\infty$ norm with the $L^p$ one, $p\geq 1$. This is partially justified by the H\"older inequality
$$\|f\|_{L^p(\Omega)}\leq|\Omega|^{1/p}\|f\|_{L^\infty(\Omega)},\quad f\in L^{\infty}(\Omega),\quad p\geq1.$$
The penalized problem reads
\begin{equation}
s^*\approx\frac{s_p^{1/p}}{|\Omega|^{1/p}},\quad s_p:=\inf_{\tilde\sbPi\in\tilde X_\Pi}\mathcal I_p(\tilde{\mbf\Pi}),\quad \mathcal I_p(\tilde{\mbf\Pi})=\int_\Omega\left[(1+\mathrm{div}\,\tilde{\mbf\Pi})^2+\ell^{-2}|\tilde{\mbf\Pi}|^2\right]^p\,dx\,.
\label{q*_p}
\end{equation}
The functional $\mathcal I_p$ is convex and twice differentiable, and finite element approximations to this problem can be found, for example, with the use of Raviart-Thomas or standard continuous piecewise-linear elements, which are conforming in the sense that the corresponding discrete space $\tilde X_{\Pi,h}$ satisfies $\tilde X_{\Pi,h}\subset \tilde X_\Pi$. Then, the discrete solution $\tilde{\mbf\Pi}_{p,h}\in \tilde X_{\Pi,h}$ defines the following lower bound of $t^*$:
\begin{equation}
t^*\geq\frac{Y}{|\hat{\mbf\sigma}_e^D|}\frac{1}{\sqrt{\mathcal I(\tilde{\mbf\Pi}_{p,h})}}.
\label{lower_bound4}
\end{equation}
It is possible also first to discretize \eqref{q*} and then to apply the penalization method to the discrete problem. 

Let us note that numerical integration is required to evaluate (\ref{q*_p}). Nevertheless, if we first discretize \eqref{q*} using, for example, lowest order Raviart-Thomas (RT0) or continuous piecewise-linear (P1) elements, then it is not necessary to carry out numerical integration. Indeed, RT0- or P1-based functions $\tilde{\mbf\Pi}_h$ are linear on any simplicial finite element $T\in \mathbb R^d$ with vertices $V_{T,1}, V_{T,1},\ldots V_{T,d+1}$. Hence,
$$\max_{x\in T}\left\{(1+\mathrm{div}\,\tilde{\mbf\Pi}_h)^2+\ell^{-2}|\tilde{\mbf\Pi}_h|^2\right\}=\max_{i=1,2,\ldots,d+1}\left\{(1+\mathrm{div}\,\tilde{\mbf\Pi}_h(V_{T,i}))^2+\ell^{-2}|\tilde{\mbf\Pi}_h(V_{T,i})|^2\right\}$$
and
$$t^*_h=\frac{Y}{|\hat{\mbf\sigma}_e^D|}\frac{1}{\sqrt{\mathcal I(\tilde{\mbf\Pi}_h)}}\leq t^*,\quad\mathcal I(\tilde{\mbf\Pi}_h)=\max_{T\in\mathcal T_h}\ \max_{i=1,\ldots,d+1}\left\{(1+\mathrm{div}\,\tilde{\mbf\Pi}_h(V_{T,i}))^2+\ell^{-2}|\tilde{\mbf\Pi}_h(V_{T,i})|^2\right\}.$$
The corresponding penalized functional can be written as
$$I_{p,h}(\tilde{\mbf\Pi}_h)=\frac{1}{p}\sum_{T\in\mathcal T_h}\ \sum_{i=1}^{d+1}\left\{(1+\mathrm{div}\,\tilde{\mbf\Pi}_h(V_{T,i}))^2+\ell^{-2}|\tilde{\mbf\Pi}_h(V_{T,i})|^2\right\}^p.$$

For numerical solution of the discrete penalized problem, we combine adaptive continuation over $p$ with the Newton method. One can expect that the larger the value of $p$, the sharper the lower bound (\ref{lower_bound4}) would be. At the same time, continuation over $p$ is important for the initialization of the Newton method. Within the continuation, we increase the values of $p$ depending on the change of the value $\mathcal I(\tilde{\mbf\Pi}_{p,h})$. This approach will be applied to examples presented in Sections \ref{lower_upper_num} and \ref{lower_upper_num_torsion}.

\subsection{Penalization method for the upper-bound problem}
\label{subsec_penalty2}

Consider the upper bound problem (see\eqref{upper_bound3})
\begin{equation}
\tilde t^*=\inf_{\substack{\tilde q\in \tilde W \\[1pt] \int_\Omega \tilde q\,dx=1}}\ \mathcal J(\tilde q),\qquad \mathcal J(\tilde q):=\int_\Omega \sqrt{\tilde q^2+\ell^2|\nabla \tilde q|^2}\,dx,
\label{t*_tilde}
\end{equation}
where
$$\tilde W=\{\tilde q\in W^{1,2}(\Omega)\ |\;\; \tilde q=0\;\mbox{on } \partial\Omega_H\}.$$
Under the assumptions from Section \ref{subsec_constant}, we have:
$$t^*=\frac{Y}{|\hat{\mbf\sigma}_e^D|}\tilde t^*,\qquad t^*\leq\frac{Y}{|\hat{\mbf\sigma}_e^D|}\mathcal J(\tilde q)\quad\forall\ \tilde q\in\tilde W.$$

The penalization described below has been successfully used in limit load analysis in perfect plasticity \cite{HRS15, HRS16, RSH18, HRS19}. The local dissipation function, that is, the integrand in (\ref{t*_tilde}), may be written in the form
\begin{equation}
D_\ell (\tilde q,\nabla\tilde q)=\sqrt{\tilde q^2+\ell^2|\nabla \tilde q|^2}=\sup_{\substack{(\tilde\pi,\tilde{\sbPi})\in \mathbb R\times\mathbb R^3\\ \tilde\pi^2+\ell^{-2}|\tilde{\sbPi}|^2\leq1}}\ \{\tilde\pi\tilde q+\tilde{\mbf\Pi}\cdot\nabla\tilde q\}.
\label{dissip}
\end{equation}
It can be penalized as follows:
\begin{equation}
D_\alpha(\tilde q,\nabla\tilde q)=\max_{\substack{(\tilde\pi,\tilde{\sbPi})\in \mathbb R\times\mathbb R^3\\ \tilde\pi^2+\ell^{-2}|\tilde{\sbPi}|^2\leq1}}\ \left\{\tilde\pi\tilde q+\tilde{\mbf\Pi}\cdot\nabla\tilde q-\frac{1}{2\alpha}(\tilde\pi^2+\ell^{-2}|\tilde{\mbf \Pi}|^2)\right\},\quad\alpha>0.
\label{dissip_alpha}
\end{equation}
We have:
\begin{equation}
D_\alpha(\tilde q,\nabla\tilde q)=\left\{
\begin{array}{ll}
\frac{\alpha}{2}(\tilde q^2+\ell^2|\nabla \tilde q|^2), & \sqrt{\tilde q^2+\ell^2|\nabla \tilde q|^2}\leq\frac{1}{\alpha}\\[2mm]
\sqrt{\tilde q^2+\ell^2|\nabla \tilde q|^2}-\frac{1}{2\alpha}, &\sqrt{\tilde q^2+\ell^2|\nabla \tilde q|^2}\geq\frac{1}{\alpha},
\end{array}
\right.
\label{dissip_alpha2}
\end{equation}
with $D_\alpha\leq D$ and $D_\alpha\rightarrow D$ as $\alpha\rightarrow+\infty$. In addition, the function $D_\alpha$ is convex, differentiable, and its second derivative exists in a generalized sense. The penalized problem reads 
\begin{equation}
\tilde t^*_\alpha=\inf_{\substack{\tilde q\in \tilde W \\[1pt] \int_\Omega \tilde q\,dx=1}}\ \mathcal J_\alpha(\tilde q),\qquad \mathcal J_\alpha(\tilde q):=\int_\Omega D_\alpha(\tilde q,\nabla\tilde q)\,dx.
\label{t*_tilde_alpha}
\end{equation}
As with the penalized lower bound problem, discrete approximations to \eqref{t*_tilde_alpha} may be sought, for example, with the use of conforming piecewise-linear or quadratic finite elements with the corresponding discrete space $\tilde{W}_h$ satisfying $\tilde{W}_h \subset \tilde{W}$.
Then, the discrete solution $\tilde{q}_{\alpha,h}^*\in \tilde W_h$ defines the following upper bound of $t^*$:
\begin{equation}
t^*\leq\frac{Y}{|\hat{\mbf\sigma}_e^D|}\mathcal J(\tilde q_{\alpha,h}^*).
\label{upper_bound4}
\end{equation}
For a convergence analysis with respect to the discretization parameter $h$, we refer to \cite{HRS15, HRS16}.

For numerical solution of the discrete counterpart to (\ref{t*_tilde_alpha}), we combine an adaptive continuation over $\alpha$ with the semismooth Newton method. One can expect that the larger the value of $\alpha$, the sharper the upper bound (\ref{upper_bound4}) would be. It is possible to study the convergence $\tilde t^*_{\alpha,h}\rightarrow\tilde t^*_h$ as $\alpha\rightarrow+\infty$, similarly as in \cite{HRS15, HRS16}. Continuation over $\alpha$ is also important for the initialization of the Newton method. Within the continuation, we increase $\alpha$ depending on the change of the value $\mathcal J(\tilde q_{\alpha,h}^*)$. This approach will be applied to examples presented in Sections \ref{lower_upper_num} and \ref{lower_upper_num_torsion}. 

\section{Example 1: plane stress problem}
\label{sec_example}

\subsection{The problem setting}
\label{subsec_setting_2D}

We consider the example introduced in \cite[Section 6.2]{CEMRS17}, of a thin plate subjected to biaxial extension. The plate is square with sides of length $L_1$ and thickness $L_2$. Its domain is thus given by $\Omega=\Omega_1\times\Omega_2\times\Omega_3$, where
$$\Omega_1=\Omega_3=(0,L_1),\quad \Omega_2=(-L_2/2,L_2/2). 
$$
The macroscopic boundary conditions are as follows (see Figure \ref{microplate}):
\begin{figure}[!h]
\centering
\includegraphics[width = 0.75\textwidth]{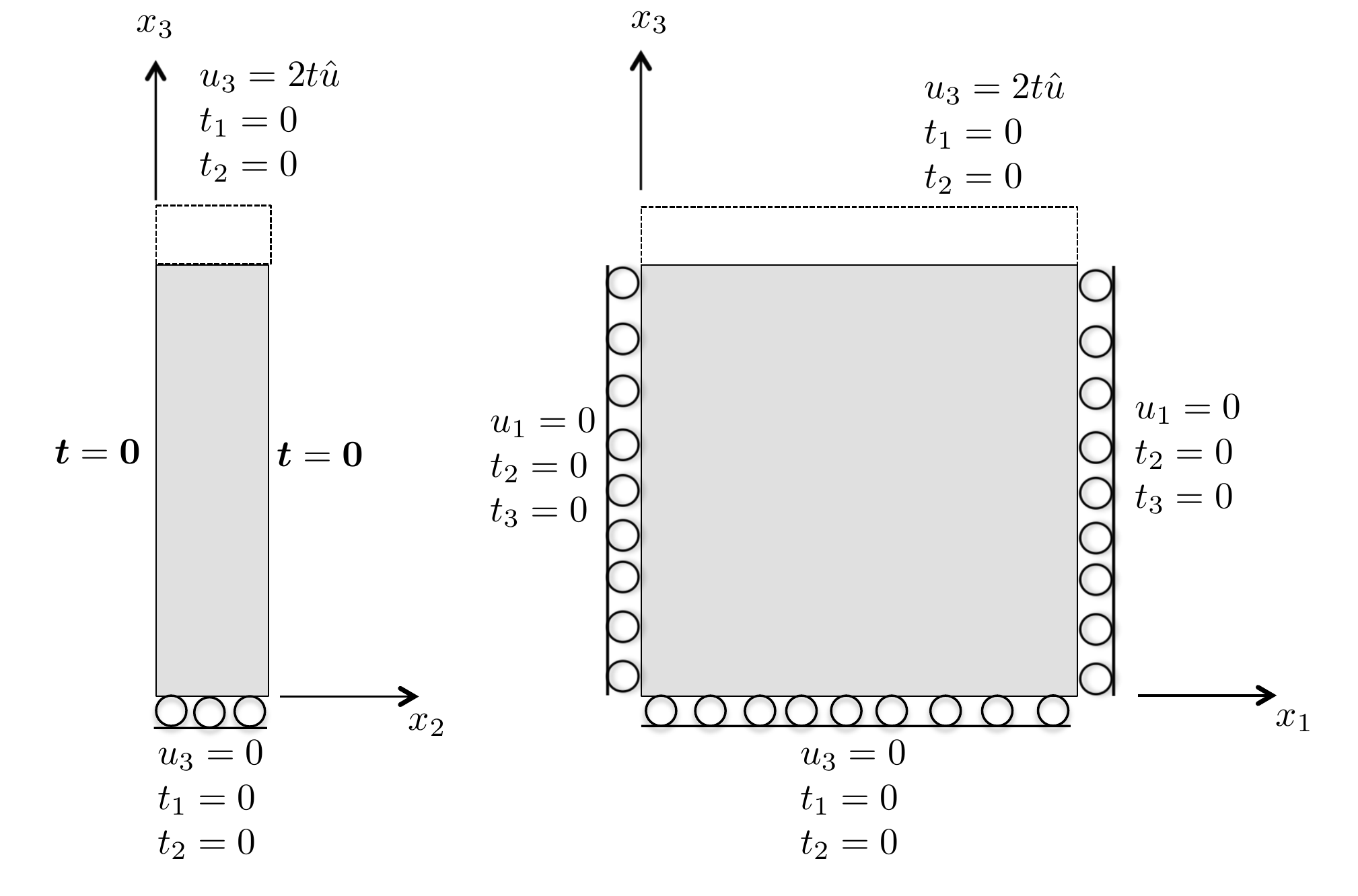}
\label{microplate}
\caption{Side and front views of the problem of a microplate subjected to uniaxial extension}
\end{figure}
    \begin{subequations}
    \begin{align}
    & t_1 = t_2 = t_3 =0 \;\;\mbox{on } \Omega_1\times\partial\Omega_2\times\Omega_3\times(0,T),
    \label{bound_y} \\
    & u_{1}= t_1 = t_3=0 \;\;\;\mbox{on } \partial\Omega_1\times\Omega_2\times\Omega_3\times(0,T),
    \label{bound_x} \\
    & u_{3}=t_1 = t_2 =0 \;\;\;\mbox{on } \Omega_1\times\Omega_2\times\{0\}\times(0,T),
    \label{bound_z0} \\
    & u_{3}=2t\hat u,\;\;t_1 = t_2 = 0 \;\;\;\mbox{on } \Omega_1\times\Omega_2\times\{L_1\}\times(0,T),
    \label{bound_z1}
    \end{align}
    \end{subequations}
where $\hat u$ is a given value, $\mbf u=(u_1,u_2,u_3)$, $\mbf t=(t_1,t_2,t_3)$ and $t\in(0,T)$. The body force is neglected, that is, $\mbf b=\mbf 0$ in $\Omega\times(0,T)$. The elastic solution (\ref{elast_branch}) can be found in closed form and is given by 
\begin{equation}
\hat{\mbf u}_e=2\frac{\hat u}{L_1}\left(\begin{array}{r}
0 \\
-\frac{\nu}{1-\nu}x_2 \\
x_3
\end{array}\right),\quad
\hat{\mbf\varepsilon}_e=2\frac{\hat u}{L_1}\left(\begin{array}{ccc}
0 & 0 & 0\\
0 & -\frac{\nu}{1-\nu} & 0\\
0 & 0 & 1
\end{array}\right),\quad 
\hat{\mbf\sigma}_e=2\lambda\frac{1-2\nu}{\nu(1-\nu)}\frac{\hat u}{L_1}\left(\begin{array}{ccc}
\nu & 0 & 0\\
0 & 0 & 0\\
0 & 0 & 1
\end{array}\right).
\label{sigma}
\end{equation}
We see that $\hat{\mbf\varepsilon}_e, \hat{\mbf\sigma}_e$ are spatially constant, so that it is possible to use the results from Section \ref{subsec_constant}. To this end we shall also need the following formulas for $\hat{\mbf\sigma}_e^D$ and its norm:
\begin{equation}
\hat{\mbf\sigma}_e^D=2\lambda\frac{1-2\nu}{3\nu(1-\nu)}\frac{\hat u}{L_1}\left(\begin{array}{ccc}
-1+2\nu & 0 & 0\\
0 & -1-\nu & 0\\
0 & 0 & 2-\nu
\end{array}\right),
\label{dev_sigma}
\end{equation}
\begin{equation}
|\hat{\mbf\sigma}_e^D|=2\lambda\sqrt{\frac{2}{3}}\frac{(1-2\nu)\sqrt{1-\nu+\nu^2}}{\nu(1-\nu)}\frac{\hat u}{L_1}.
\label{norm_dev_sigma}
\end{equation}

Next, we need to define microscopic boundary conditions. To be consistent with the plane stress assumptions, we impose the micro-free conditions 
\begin{equation}
\Pi_{ij2} = 0\;\;\mbox{on } \Omega_1\times\partial\Omega_2\times\Omega_3\times(0,T)\,.
\label{bound_y2}
\end{equation}
On the remaining parts of the boundary we consider either micro-free or micro-hard boundary conditions; that is, 
\begin{equation}
\Pi_{ij1} = 0\;\;\mbox{on } \partial\Omega_1\times\Omega_2\times\Omega_3\times(0,T),\quad \Pi_{ij3} = 0\;\;\mbox{on } \Omega_1\times\Omega_2\times\partial\Omega_3\times(0,T),
\label{bound_micro-free}
\end{equation}
or
\begin{equation}
\mbf\varepsilon^p=\mbf 0\;\;\;\mbox{on } [\partial\Omega_1\times\Omega_2\times\Omega_3\times(0,T)]\cup[\Omega_1\times\Omega_2\times\partial\Omega_3\times(0,T)].
\label{micro_hard}
\end{equation}

In the following subsections, we derive lower and upper bounds of the elastic threshold $t^*$ based on analytical and numerical approaches. To this end, we use the following input data:
\begin{equation}
L_1=50, \; L_2=2,\; T=1/2,\; \lambda=1.05\times10^{-1},\; \nu=0.3,\; \hat u=2/3,\; Y=1\times10^{-3},\; \ell=5.
\label{data}
\end{equation}

\subsection{Lower and upper bounds of $t^*$ based on analytical approach}
\label{subsec_anal_2D}

If we consider the micro-free boundary condition \eqref{bound_micro-free}, the spaces $\tilde X_\Pi$ and $\tilde W$ corresponding to the estimates (\ref{lower_bound3}) and (\ref{upper_bound3}) are 
$$\tilde X_\Pi=\{\tilde{\mbf\Pi}\in L^\infty(\Omega;\mathbb R^3)\;|\;\; \mathrm{div}\,\tilde{\mbf \Pi}\in L^\infty(\Omega;\mathbb R^3),\;\; \tilde{\Pi}_{ijk}=0\;\mbox{ for }\ x_k\in\partial\Omega_k,\; k=1,2,3\},$$
$$\tilde W= W^{1,2}(\Omega).$$
Choosing $\tilde{\mbf\Pi}=\mbf 0$ and $\tilde q=1$ in (\ref{lower_bound3}) and (\ref{upper_bound3}), respectively, we find that the lower and upper bounds coincide and thus
\begin{equation}
t^*=t^*_L=t^*_U=\frac{Y}{|\hat{\mbf\sigma}_e^D|}=\sqrt{\frac{3}{2}}\frac{YL_1\nu(1-\nu)}{2\hat u\lambda(1-2\nu)\sqrt{1-\nu+\nu^2}}\doteq 0.2584.
\label{t^*_ex1}
\end{equation}
This coincides also with the solution for the case of conventional plasticity. To complete the elastic solution from (\ref{elast_branch}) for any $t\in [0,t^*]$, we set $\hat{\mbf\Pi}_e=\mbf0$ and $\hat{\mbf\pi}_e=\hat{\mbf\sigma}_e^D$.

If we consider the micro-hard boundary condition \eqref{micro_hard}, the spaces $\tilde X_\Pi$ and $\tilde W$ corresponding to the estimates (\ref{lower_bound3}) and (\ref{upper_bound3}) are now
$$\tilde X_\Pi=\{\tilde{\mbf\Pi}\in L^\infty(\Omega;\mathbb R^3)\;|\;\; \mathrm{div}\,\tilde{\mbf \Pi}\in L^\infty(\Omega;\mathbb R^3),\;\; \tilde{\Pi}_{ij2}=0\;\mbox{ for }\ x_2\in\partial\Omega_2\},$$
$$\tilde W=\{\tilde q\in W^{1,2}(\Omega)\ |\;\; \tilde q=0\;\mbox{ for }\ x_i\in\partial\Omega_i,\;\; i=1,3\}.$$
Unlike the micro-free boundary condition, we see that $\tilde q=1\not\in\tilde W$. Therefore, (\ref{t^*_ex1}) is now only a lower bound. To derive a sharper lower bound than $t^*_{L,1}\doteq 0.2584$, we consider $\tilde{\mbf \Pi}$ in the form (see \eqref{piPiq})
\begin{equation}
\tilde\Pi_{i}(\mbf x)=a_{i}x_i+b_{i},\quad a_{i}, b_{i}\in\mathbb R,\;\; i=1,3,\ \mbox{no sum on}\ i\,.
\label{Pi_lin}
\end{equation}


We shall optimize the parameters $a_i$, $b_i$ to achieve the largest value of (\ref{lower_bound3}). To this end, we minimize the functional  
\begin{equation}
\mathcal I(\tilde{\mbf\Pi}):=\left\|(1+\mathrm{div}\,\tilde{\mbf\Pi})^2+\ell^{-2}|\tilde{\mbf\Pi}|^2\right\|_{L^\infty(\Omega)}=\max_{\sbx\in\Omega}\left\{(1+\mathrm{div}\,\tilde{\mbf\Pi})^2+\ell^{-2}|\tilde{\mbf\Pi}|^2\right\}
\label{cost_function1}
\end{equation}
representing the denominator in (\ref{lower_bound3}). 
The maximum over $x_i\in\Omega_i$ is achieved at one of the end points of the interval $\Omega_i$, i.e., for $x_i\in\partial\Omega_i$, $i=1,3$. To minimize the terms $|a_{i}x_i+b_{i}|$ on $\partial\Omega_i$, we set
$b_{i}=-\frac{L_1}{2}a_{i}\ (i=1,3)$, which
implies
\begin{equation}
\mathcal I(\tilde{\mbf\Pi})=(1+a_1+a_3)^2+\left(\frac{L_1}{2\ell}\right)^{2}(a_{1}^2+a_{3}^2).
\label{criterion2}
\end{equation}
The optimal values of $a_{i}$ are found by minimization of (\ref{criterion2}), leading to
the following lower bound of $t^*$:
\begin{equation}
t_{L,2}^*=\frac{Y}{|\hat{\mbf\sigma}_e^D|}\frac{2\ell}{L_1}\sqrt{2+\frac{L_1^2}{4\ell^2}}\doteq 0.2685.
\label{threshold}
\end{equation}

We see that the elastic threshold $t^*_{L,1}$ for perfect (Hencky) plasticity is smaller than $t_{L,2}^*$, i.e., then $t^*$. Next, from the presented results, it follows that the microstress $\mbf\Pi$ corresponding to the solution to the gradient plasticity problem is not unique. Especially for sufficiently small $t>0$, we derived two different solutions corresponding to (\ref{elast_branch}): $\mbf\Pi=\mbf 0$ and $\mbf\Pi=t\hat{\mbf \sigma}_e^D\otimes\tilde{\mbf\Pi}$, where
\begin{equation*}
\tilde\Pi_{1}(\mbf x)=\frac{\frac{L_1}{2}-x_1}{2+\frac{L_1^2}{4\ell^2}},\quad \tilde\Pi_{2}(\mbf x)=0,\quad \tilde\Pi_{3}(\mbf x)=\frac{\frac{L_1}{2}-x_3}{2+\frac{L_1^2}{4\ell^2}}.
\end{equation*}

Now, we estimate $t^*$ from above. We choose $\tilde q(\mbf x)=x_1(L_1-x_1)x_3(L_1-x_3)\in\tilde W$. Then (\ref{upper_bound3}) yields $t^*_{U}\doteq 0.3108$ using numerical integration. This bound may be compared with the value of approximately 0.34 for the elastic threshold, as seen in \cite[Figure 8]{CEMRS17}. That value corresponds to the threshold at a particular point in the domain, and it is clear from the bounds obtained here that the elastic threshold will have been reached earlier than determined from this arbitrary point in the domain. 

There is a gap between the upper and lower bounds that may be narrowed by optimizing $\tilde{\mbf\Pi}$ and $\tilde q$ through a numerical procedure.

\subsection{Lower and upper bounds of $t^*$ based on numerical methods}
\label{lower_upper_num}

We consider the micro-hard boundary conditions \eqref{micro_hard}. Under the plane stress assumption, the lower and upper bound problems are posed on the domain ${\tilde\Omega}=(0,50)\times(0,50)$. In particular, the lower bound problem reads
\begin{equation}
t^*=\frac{Y}{|\hat{\mbf\sigma}_e^D|}\frac{1}{\sqrt{\inf\limits_{\tilde\sbPi\in\tilde X_\Pi}\mathcal I(\tilde{\mbf\Pi})}},\qquad \mathcal I(\tilde{\mbf\Pi})=\left\|(1+\mathrm{div}\,\tilde{\mbf\Pi})^2+\ell^{-2}|\tilde{\mbf\Pi}|^2\right\|_{L^\infty(\tilde\Omega)},
\label{q*2}
\end{equation}
where $\tilde X_\Pi=\{\tilde{\mbf\Pi}\in L^\infty(\tilde\Omega;\mathbb R^2)\;|\;\; \mathrm{div}\,\tilde{\mbf \Pi}\in L^\infty(\tilde\Omega;\mathbb R^2)\},$ and the upper bound one reduces to
\begin{equation}
t^*=\frac{Y}{|\hat{\mbf \sigma}_e^D|}\tilde t^*,\qquad \tilde t^*=\inf_{\substack{\tilde q\in \tilde W \\[1pt] \int_{\tilde\Omega} \tilde q\,dx=1}}\ \int_{\tilde\Omega} \sqrt{\tilde q^2+\ell^2|\nabla \tilde q|^2}\,dx,\quad \tilde W=W^{1,2}_0(\tilde\Omega).
\label{t*U2}
\end{equation}
We also know that ${Y}/{|\hat{\mbf \sigma}_e^D|}\doteq 0.2584$.

To find lower and upper bounds of $t^*$ based on numerical techniques, we use the penalization methods presented in Section \ref{sec_penalty}. For the solution of both problems, we use a regular mesh with $401\times 401$ nodes. Problem \eqref{q*2} is discretized using RT0 elements, while piecewise quadratic P2 elements with 7-point Gauss quadrature are used for solving the problem \eqref{t*U2}. Both problems are implemented in MATLAB.

\begin{figure}[!h]
	\centering
	\includegraphics[width = 0.45\textwidth]{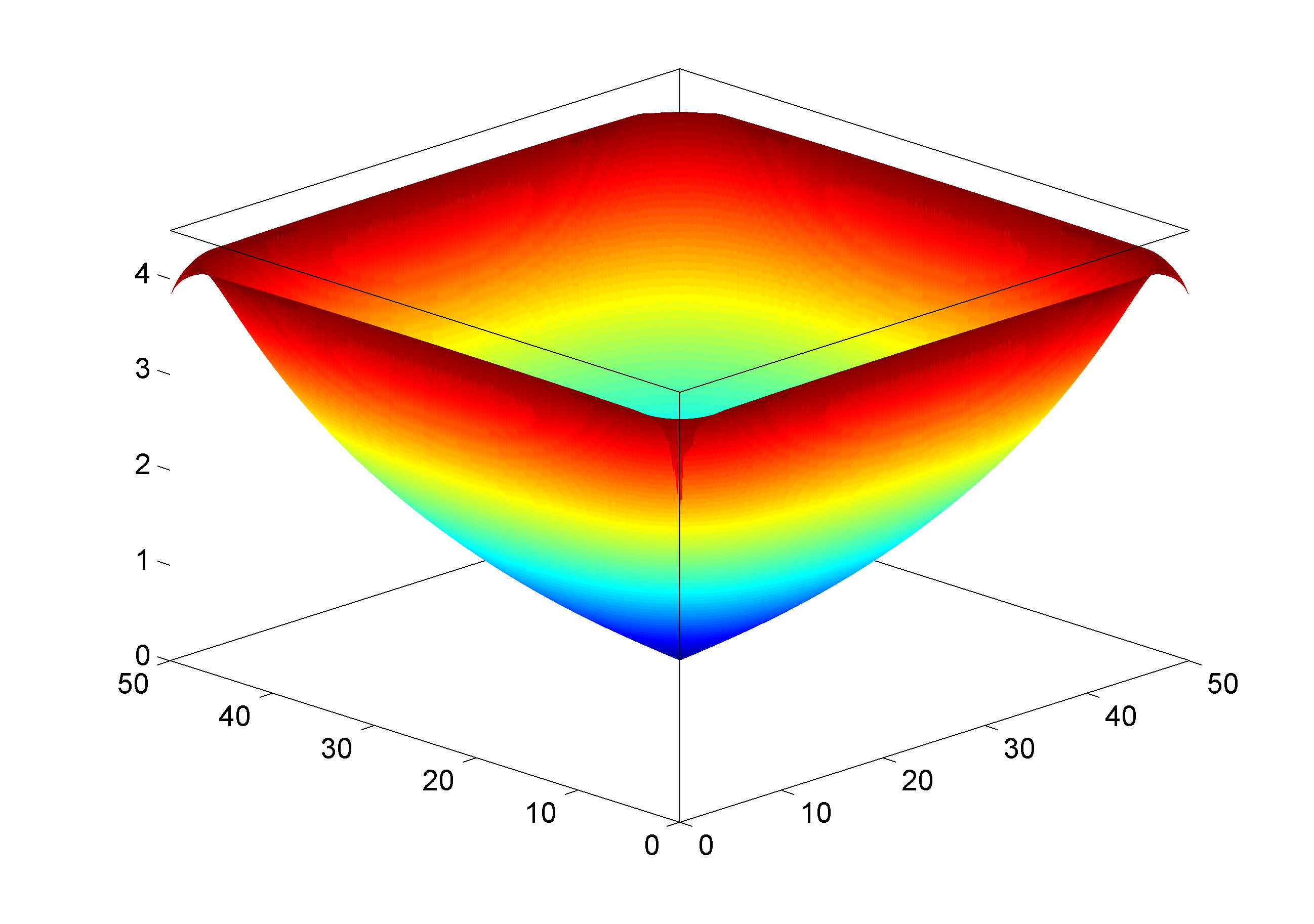}
	\includegraphics[width = 0.45\textwidth]{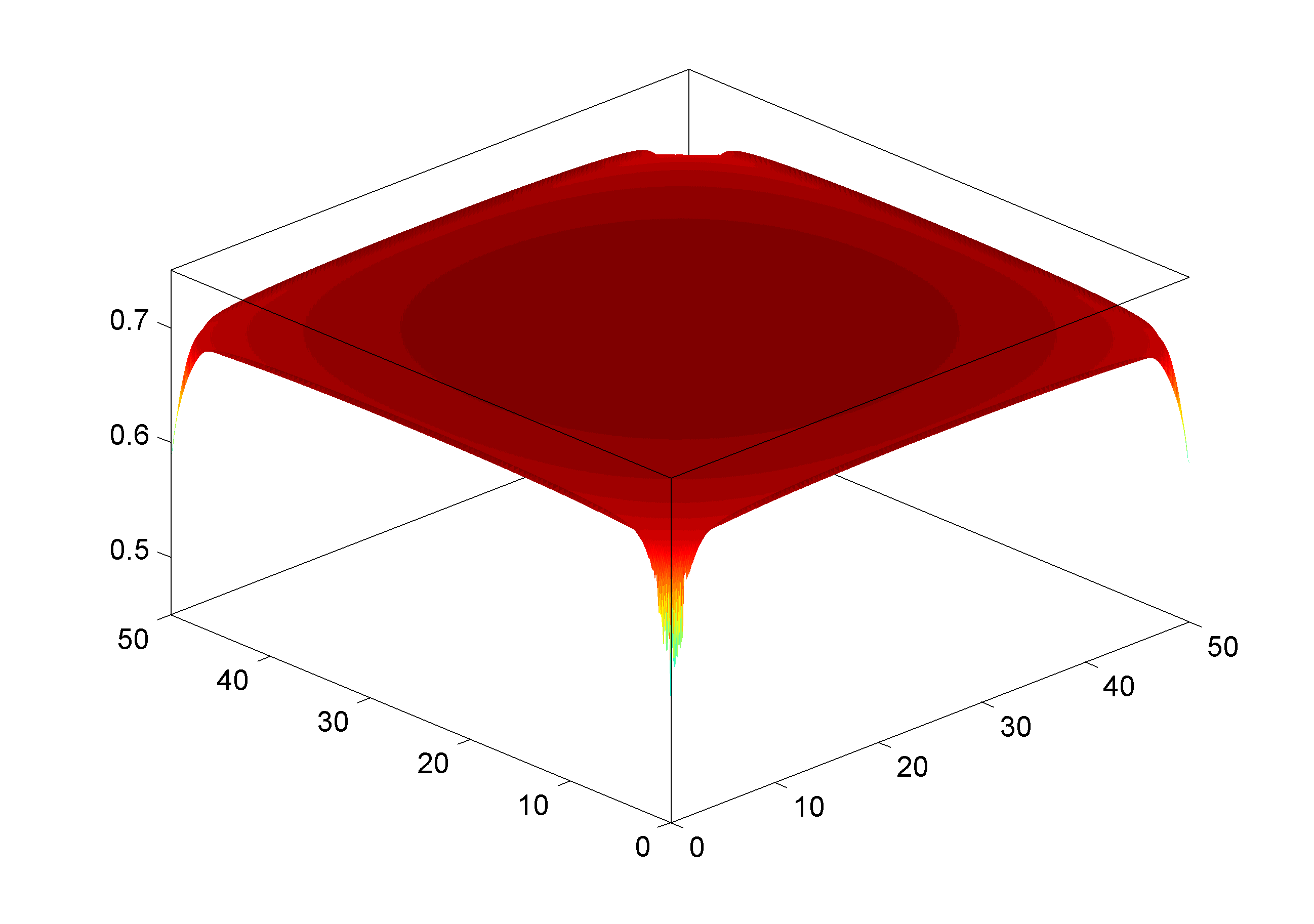}\\
	(a) \hspace{36ex} (b) \\
	\caption{(a) The norm $|\tilde{\mbf\Pi}|$ of the solution to \eqref{q*2}; (b) the corresponding functional $\mathcal I$}
	\label{fig_Pi_plate}
\end{figure}

\begin{figure}[!h]
	\centering
	\includegraphics[width = 0.45\textwidth]{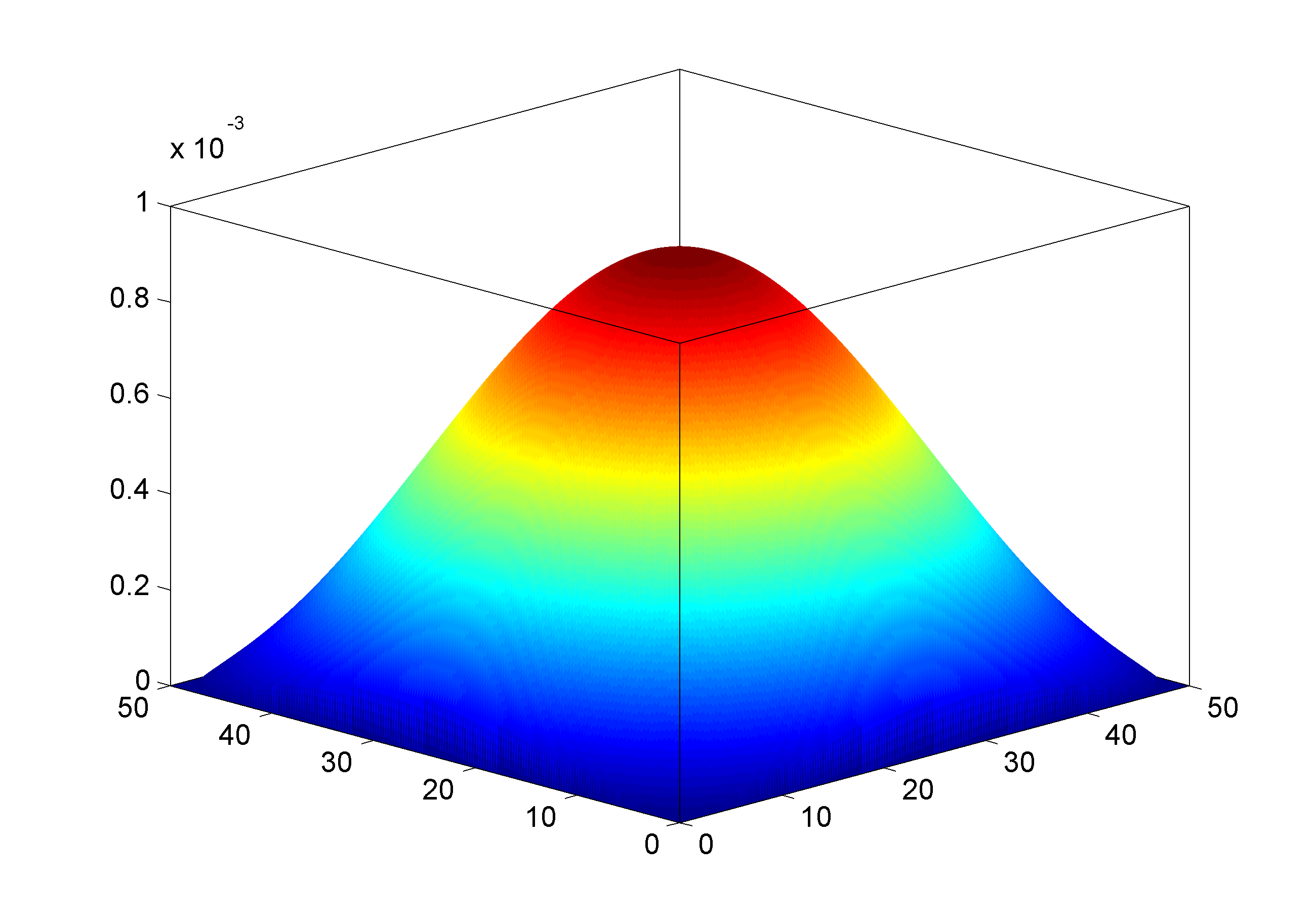}
	\includegraphics[width = 0.45\textwidth]{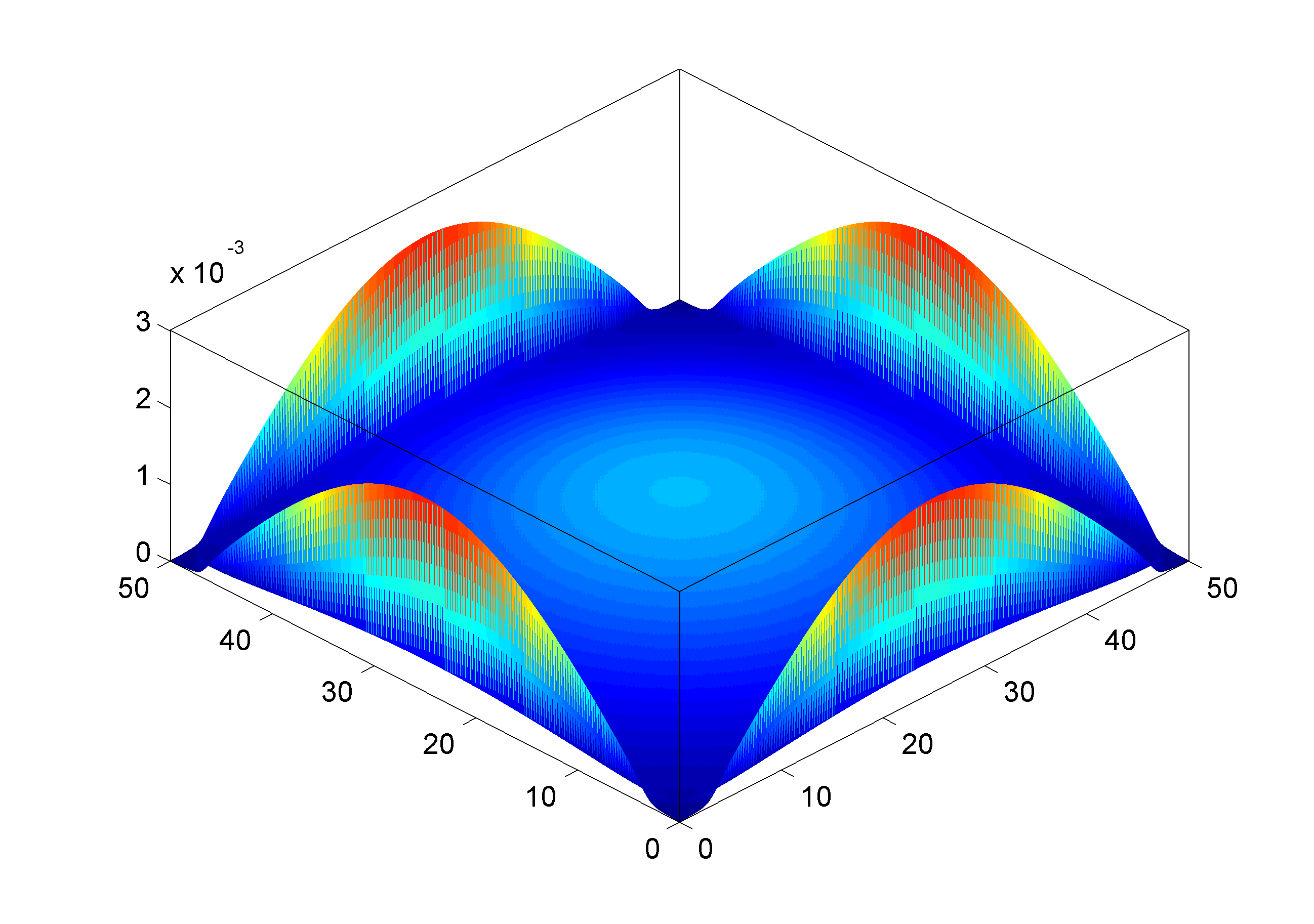}\\
	(a) \hspace{36ex} (b) \\	
	\caption{(a) The solution $\tilde q$ to \eqref{t*U2}; (b) the corresponding local dissipation $D_\ell (\tilde q )$}
	\label{fig_q_plate}
\end{figure}

\begin{figure}[!h]
	\centering
	\includegraphics[width = 0.45\textwidth]{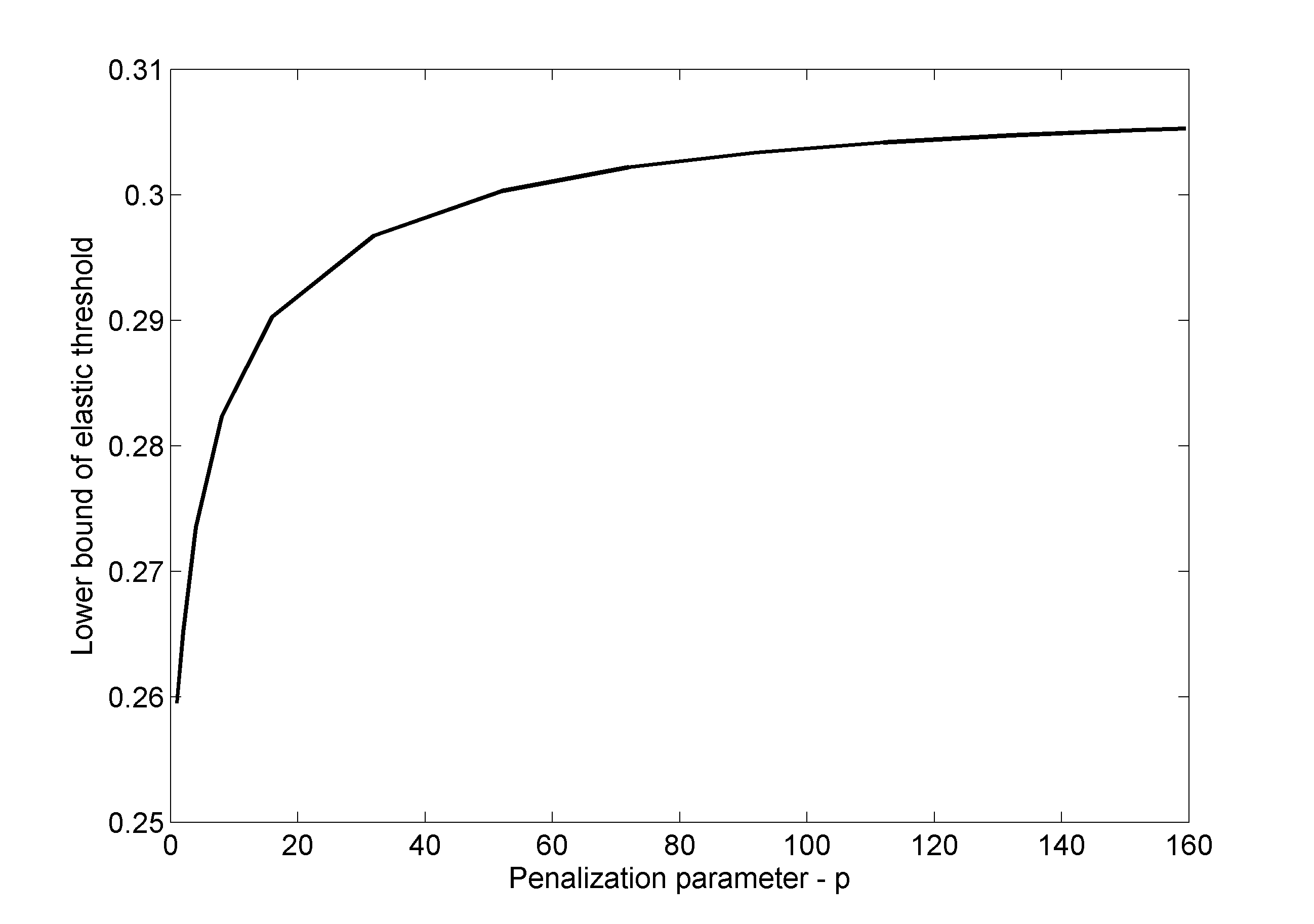}
	\includegraphics[width = 0.45\textwidth]{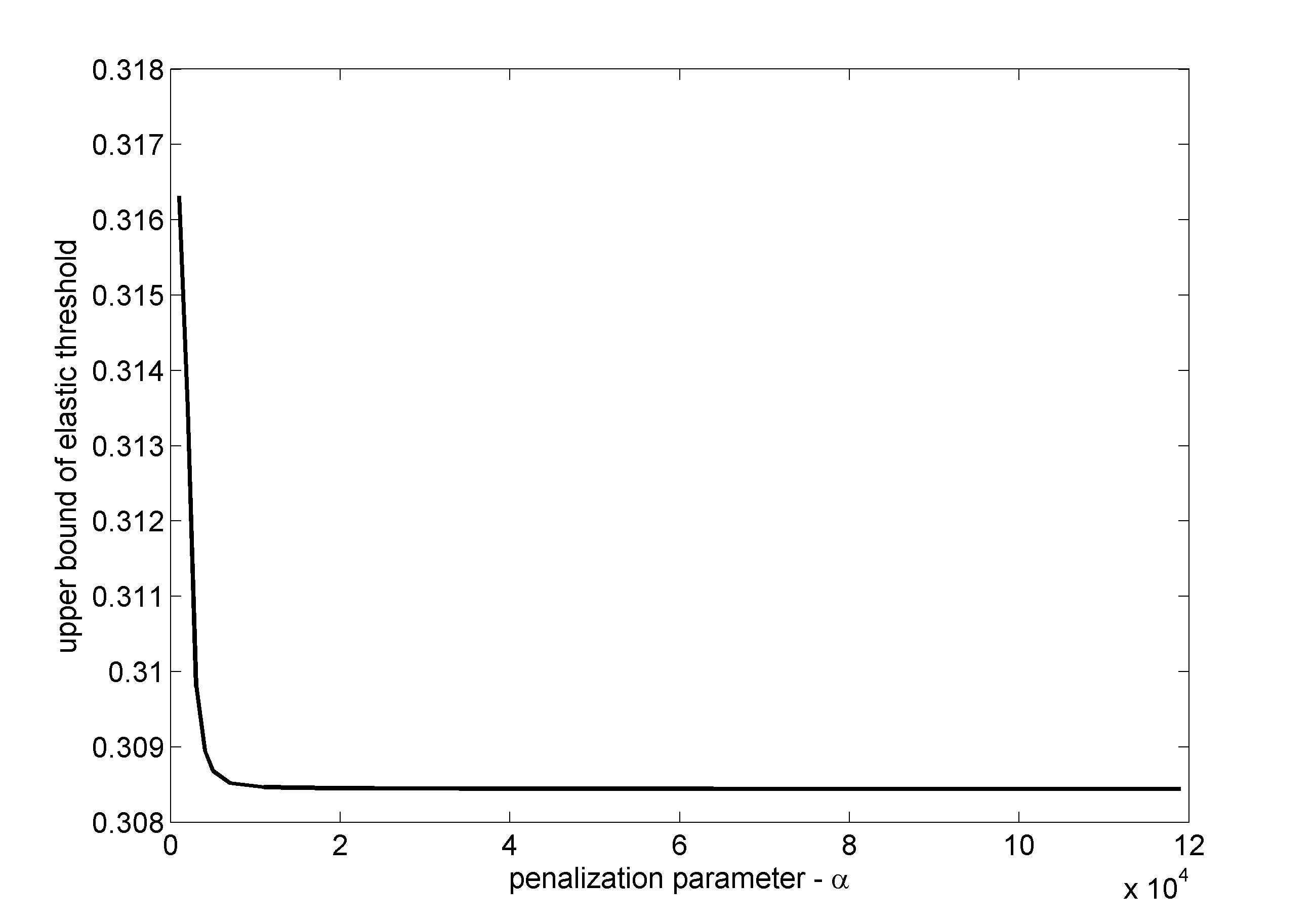}\\
	(a) \hspace{36ex} (b) \\
	\caption{Lower (a) and upper (b) bounds $t^*$ corresponding to the functionals \eqref{q*2} and \eqref{t*U2}}
	\label{fig_bounds_plate}
\end{figure}

Computed lower and upper bounds depending on the penalization parameters $p$ and $\alpha$, respectively, are depicted in Figure \ref{fig_bounds_plate}. From the largest used values of these parameters, we find the following bounds of $t^*$:
$$0.3052\leq t^*\leq 0.3085.$$
We see that both the lower and upper bounds are very sharp. In particular, the lower bound has been improved in comparison of Section \ref{subsec_anal_2D}.

The numerical solution to the problem \eqref{q*2} arising from the largest value of $p$ is depicted in Figure \ref{fig_Pi_plate}. In particular, the norm of $|\tilde{\mbf \Pi}|$ is visualized on the left figure, while local values defining the functional $\mathcal I$ are depicted on the right. This solution does not have a simple shape, with a corner effect evident. From the right figure, one can see that the local values of $\mathcal I$ are almost constant except at the corner points. Similar results have been obtained with the use of P1 elements.

The numerical solution to the problem \eqref{t*U2} corresponding to the largest value of $\alpha$ is depicted in Figure \ref{fig_q_plate}. Also shown is the corresponding local dissipation $D_\ell$ (see \eqref{dissip}). This figure shows large values of $\nabla \tilde q$ along the edges.
%

%
 
\section{Example 2: torsion of a micro-rod}
\label{sec_torsion}

\subsection{The problem setting}

Consider a rod of radius $R$ and length $L$, with $0 < L \leq \infty$, subject to a uniform torsion with $\kappa$ denoting the twist per unit length. The only non-zero displacement, relative to a cylindrical coordinate system $(r,\vartheta,z)$ is $u_{\vartheta} = \kappa zr$. The only stress in the elastic region is then $\sigma_{\vartheta z}= \mu\kappa r$, where $\mu$ is the shear modulus. In accordance with the notation introduced in Section \ref{sec_threshold}, we set $\kappa(t)=t\hat\kappa$, where $\hat\kappa>0$ is a given value. Then it follows that $\hat{\mbf\sigma}^D_e=\hat{\mbf\sigma}_e$ and $|\hat{\mbf\sigma}^D_e|=\mu\hat\kappa r$. We shall also use the notation $\hat\sigma_{\vartheta z}= \mu\hat\kappa r$.

In the context of size-dependent and microscale mechanical response, this problem has been studied experimentally \cite{Fleck_etal1994} as well as theoretically and computationally. The study \cite{Idiart-Fleck2010} shows the link between strengthening, that is, an increase in yield strength with length scale, which characterizes the dissipative problem, the topic of this work. The work \cite{Chiricotto_etal2016a} studies the energetic problem, in which gradient terms arise through a recoverable energy. In this case the link is between hardening, that is, an increase in slope in the plastic range, and increase in length scale. In \cite{Bardella-Panteghini2015} the torsion problem is discussed in the context of a more elaborate gradient theory that takes account of plastic rotation and dislocation density. 

We assume that $\Pi_{\vartheta z r}(r)$ and $\Pi_{r z \vartheta}(r)$ are the only non-zero components of $\bPi$. Then microforce balance \eqref{microforce_balance} and the yield function \eqref{yield} reduce respectively to
\begin{equation}
\sigma_{\vartheta z} =\pi_{\vartheta z} - \Pi_{\vartheta zr,r} - r^{-1}(\Pi_{\vartheta z r} + \Pi_{r z \vartheta}) 
\label{mfbpolar2}
\end{equation}
and 
\begin{equation}
f_\ell(\pi_{\vartheta z},\Pi_{\vartheta zr},\Pi_{r z \vartheta})=\sqrt{\pi_{\vartheta z}^2+\ell^{-2}(\Pi_{\vartheta zr}^2 + \Pi_{r z \vartheta}^2)}-Y\,.
\label{yield_reduce}
\end{equation}
in $(0,R)$. The boundary conditions are a traction-free surface $r = R$, which is automatically satisfied, a microhard condition at $r = 0$, and a microfree condition at $r = R$. That is,
\begin{equation}
u_\vartheta (0) = 0,\qquad \varepsilon^p_{\vartheta  z}(0) = 0,\qquad \ \Pi_{\vartheta z r}(R) = 0\,.
\label{bcs_torsion}
\end{equation}

From \eqref{t*0}, \eqref{mfbpolar2} and \eqref{yield_reduce}, the elastic threshold for this problem is given by
\begin{align}
t^*& =\max\Big\{t>0\ |\;\; \exists(\pi,\Pi,\bar\Pi)\in X:\;\; t\hat\sigma= \pi-\Pi'- r^{-1}(\Pi + \bar\Pi),
\nonumber \\
& \qquad\qquad\qquad \sqrt{\pi^2+\ell^{-2}(\Pi^2 + \bar\Pi^2)}\leq Y\;\;\mbox{in }(0,R), \ \Pi(R) = 0\Big\},
\label{t*5}
\end{align}
where $\hat\sigma:=\hat\sigma_{\vartheta z}$, $\Pi: = \Pi_{\vartheta z r}$ and $\bar{\Pi} := \Pi_{r z \vartheta }$.
If $Y$ is constant, then from \eqref{lower_bound1} we have
\begin{equation}
t^*=\frac{Y}{\min\limits_{(\Pi,\bar\Pi)\in X_\Pi}\max\limits_{r\in[0,R]}\sqrt{(\hat\sigma+
\Pi' + r^{-1}(\Pi+\bar\Pi))^2
+\ell^{-2}(\Pi^2 + \bar\Pi^2)}},
\label{lower_bound8}
\end{equation}
where
\begin{align}
X_\Pi &=\big\{(\Pi,\bar\Pi)\in W^{1,\infty}((0,R))\times L^{\infty}((0,R))\ |\;\; \Pi(R)=0,\nonumber\\
& \qquad\qquad r^{-1}(\Pi+\bar\Pi)\in L^{\infty}((0,R))\big\}.
\label{X_Pi}
\end{align}
We may render \eqref{lower_bound8} dimensionless by setting
\[
\tilde r:=\frac{r}{R},\ \ \tilde \ell=\frac{\ell}{R},\ \  \tilde{\Pi} = \frac{\Pi}{\mu \hat{\kappa}R^2},\ \  \tilde{\bar{\Pi}} = \frac{\Pi}{\mu \hat{\kappa}R^2}\,.
\]
Then \eqref{lower_bound8} becomes
\begin{equation}
t^*=\frac{Y}{\mu\hat{\kappa} R}\tilde t^*,\quad \tilde t^*=\frac{1}{\sqrt{\min\limits_{(\tilde \Pi,\tilde{\bar\Pi})\in \tilde X_\Pi} \mathcal I(\tilde \Pi,\tilde{\bar\Pi})}}
\label{lower_bound5},
\end{equation}
with 
\begin{equation}
\mathcal I(\tilde \Pi,\tilde{\bar\Pi})=\max\limits_{\tilde r\in[0,1]}\left[\left(
\tilde r + \tilde\Pi'(\tilde r) + {\tilde r}^{-1}(\tilde \Pi(\tilde r)+\tilde{\bar\Pi}(\tilde r))\right)^2
	+\tilde\ell^{-2}\left(\tilde{\Pi}^2(\tilde r) + \tilde{\bar\Pi}^2(\tilde r)\right)\right]
\label{I_Pi}
\end{equation}
and
\begin{align}
\tilde X_\Pi &=\big\{(\tilde\Pi,\tilde{\bar\Pi})\in W^{1,\infty}((0,1))\times L^{\infty}((0,1))\ |\;\; \tilde\Pi(1)=0,\nonumber\\
& \qquad\qquad r^{-1}(\tilde\Pi+\tilde{\bar\Pi})\in L^{\infty}((0,1))\big\}.
\label{X_Pi2}
\end{align}
We see that $\tilde t^*$ depends only on $\tilde\ell$. In \eqref{lower_bound5}--\eqref{X_Pi2}, it is also convenient to introduce $\tilde{\hat\Pi}:=r^{-1}(\tilde\Pi+\tilde{\bar\Pi})$ and to eliminate $\tilde{\bar\Pi}$ in \eqref{I_Pi} by the substitution 
\begin{equation}
\tilde{\bar\Pi}=r\tilde{\hat\Pi}-\tilde\Pi\,.
\label{Pihat}
\end{equation}
It should be noted that one cannot then use the weighted functional space; see Section \ref{subsec_discrete_torsion}.

To derive the dual definition of the threshold $t^*$, we introduce a weak form for the microforce balance equation \eqref{mfbpolar2}. 
Given the homogeneous Dirichlet condition \eqref{bcs_torsion}, we define
\begin{equation}
W = \{ q \in W^{1,2}(0,R)\ |\ q(0) = 0\}\,.
\label{W}
\end{equation}
After multiplying \eqref{mfbpolar2} by $q rdr$ and integrating we obtain
\[
t\int_0^R \hat{\sigma}qr\ dr = \int_0^R [ \pi qr + \Pi(rq') - \bar{\Pi}q] \ dr \quad\forall q\in W\,.
\]
The expression for $t^*$ then becomes
\begin{align}
t^*& =\sup_{\substack{(\pi,\Pi,\bar{\Pi})\in X\\ \sqrt{\pi^2+\ell^{-2}(\Pi^2 + \bar\Pi^2)}\leq Y\;\mbox{{\scriptsize in }}(0,R)}}\;\;\inf_{\substack{q\in W\\ \int_0^R\hat\sigma qr\,dr=1}}\;\int_0^R  [ \pi qr + \Pi(rq') - \bar{\Pi}q] \,dr\nonumber\\
&=\inf_{\substack{q\in W\\ \int_0^R\hat\sigma qr\,dr=1}}\;\int_0^R\sup_{\substack{(\pi,\Pi)\in X\\ \sqrt{\pi^2+\ell^{-2}(\Pi^2 + \bar{\Pi}^2)}\leq Y}}
[ \pi qr + \Pi(rq') - \bar{\Pi}q] \,dr\nonumber\\
&=\inf_{\substack{q\in W\\ \int_0^R\hat\sigma qr\,dr=1}}\;\int_0^R Y\sqrt{q^2+\ell^2[(q')^2 + (q/r)^2]}\, r\,dr.
\label{lower_bound7}
\end{align}

Substituting $\hat\sigma=\mu\hat\kappa r$, $\tilde r:=r/R$, $\tilde \ell=\ell/R$, \eqref{lower_bound7} can be transformed to
\begin{equation}
t^*=\frac{Y}{\nu\kappa R}\tilde t^*,\quad \tilde t^*=\inf_{\substack{q\in \tilde W\\ \int_0^1 \tilde r^2q\,d\tilde r=1}}\;\int_0^1 \sqrt{q^2+\tilde\ell^2[(q')^2 + (q/\tilde r)^2]}\, \tilde r\,d\tilde r,
\label{lower_bound9}
\end{equation}
where $\tilde W = \{ q \in W^{1,2}(0,1)\ |\; q(0) = 0\}$.

\subsection{Upper and lower bounds of $\tilde t^*$ based on an analytical approach}
\label{subsec_bounds_anal_torsion}

To find an upper bound of $\tilde t^*$ we choose $q(\tilde r) = 4\tilde r$, in \eqref{lower_bound9}. This yields
\begin{equation}
\tilde t^*\leq\tilde t^*_U = \frac{4}{3}\left\{\left[ 1 + 2\tilde\ell^2 \right]^{3/2}-(2\tilde\ell^2)^{3/2}\right\}\leq \frac{4}{3}\left[ 1 + 2\tilde\ell^2 \right]^{3/2}\,.
\label{upper_bound_torsion}
\end{equation}
If $\tilde\ell=0$ then choosing $q(\tilde r) = (n+1)r^n$, we obtain $t^*_U\rightarrow 1$ as $n\rightarrow+\infty$.

To determine a lower bound consider first of all $\tilde\Pi=\tilde{\bar\Pi}= 0$. This gives, from \eqref{lower_bound5},
\begin{equation}
\tilde t^*\geq \tilde t^*_{L,1} = 1\,.
\label{lower_bound1_torsion}
\end{equation}
We see that $\tilde t^*_{L,1}=\tilde t^*=1$ for $\tilde\ell=0$. To obtain a sharper lower bound one can choose
	\begin{equation}
	\tilde\Pi=a\tilde r(\tilde r-1), \quad \tilde{\bar\Pi}=0,\quad \tilde r\in[0,1],
	\label{Pi_choice}
	\end{equation}
	where the parameter $a\in\mathbb R$ is optimized. This bound is depicted and compared against other estimates in Figure \ref{fig_torsion}.

\subsection{Discretization and penalization of the lower bound problem} 
\label{subsec_discrete_torsion}

We adopt the following version of the lower bound problem, from \eqref{I_Pi} and \eqref{Pihat} (tilde symbols are omitted for the sake of simplicity):
\begin{equation}
t^*=\frac{Y}{\mu\hat{\kappa} R}\tilde t^*,\quad \tilde t^*=\frac{1}{\sqrt{\min\limits_{(\Pi,\hat\Pi)\in \hat X_\Pi} \mathcal I(\Pi,\hat\Pi)}}
\label{lower_bound10},
\end{equation}
where $\hat X_\Pi =\big\{(\Pi,\hat\Pi)\in W^{1,\infty}((0,1))\times L^{\infty}((0,1))\ |\;\; \Pi(1)=0\big\}$ and
\begin{equation}
\mathcal I(\Pi,\hat\Pi)=\max\limits_{r\in[0,1]}\left\{\left(
r + \Pi'(r) + \hat\Pi(r))\right)^2
+\tilde\ell^{-2}\left[\Pi^2(r) + \left(r\hat\Pi(r)-\Pi(r)\right)^2\right]\right\}\,.
\label{I_Pi3}
\end{equation}
We discretize and penalize this problem as in Section \ref{subsec_penalty1}. Consider
a uniform partition of the interval $[0,1]$:
$$0=r_1<r_2<\ldots<r_{N+1}=1,\quad r_{i+1}-r_{i}=h=1/N,\;\; i=1,2,\ldots,N\,,$$
and the following approximation of $\hat X_\Pi$:
\begin{align*}
\hat X_{\Pi,h} &=\big\{(\Pi_h,\hat\Pi_h)\in C([0,1])\times L^{\infty}((0,1))\ \big|\; \Pi_h(1)=0,\\
&\qquad\qquad \Pi_h|_{(r_{i},r_{i+1})}\in P_1,\;\; 
\hat{\Pi}_h|_{(r_{i},r_{i+1}})\in P_0,
\;\; i=1,2,\ldots,N\big\}\,.
\end{align*}
Here $P_1$ and $P_0$ are respectively the spaces of polynomials of degree 1, and of constants, on the interval $[r_i,r_{i+1}]$.
Clearly, $\hat X_{\Pi,h}\subset\hat X_{\Pi}$, which implies 
\begin{equation}
t^*\geq t^*_{L,h}=\frac{Y}{\mu\hat{\kappa} R}\tilde t^*_{L,h},\quad \tilde t^*_{L,h}=\frac{1}{\sqrt{\min\limits_{(\Pi_h,\hat\Pi_h)\in \hat X_{\Pi,h}} \mathcal I(\Pi_h,\hat\Pi_h)}}\,.
\label{lower_bound11}
\end{equation}
In addition, we have
\begin{align}
\mathcal I(\Pi_h,\hat\Pi_h) &=\max\limits_{i}\ \max\limits_{r\in[r_{i},r_{i+1}]}\left\{\left(
r + \Pi'_h + \hat\Pi_h)\right)^2
+\tilde\ell^{-2}\left[\Pi^2_h + \left(r\hat\Pi_h-\Pi_h\right)^2\right]\right\}\nonumber\\
&=\max\limits_{i}\ \max\Big\{[
r_{i} + \Pi'_{h,i} + \hat\Pi_{h,i}]^2
+\tilde\ell^{-2}[\Pi^2_h(r_{i}) + (r_{i}\hat\Pi_{h,i}-\Pi_h(r_{i}))^2],\nonumber\\
&\qquad\qquad\qquad\;\; [r_{i+1} + \Pi'_{h,i} + \hat\Pi_{h,i}]^2
+\tilde\ell^{-2}[\Pi^2_h(r_{i+1}) + (r_{i+1}\hat\Pi_{h,i}-\Pi_h(r_{i+1}))^2]\Big\}\,.
\label{I_Pi4}
\end{align}
Here we have used the property that the quadratic function defined in $[r_{i},r_{i+1}]$ has a maximum either at $r_{i}$ or $r_{i+1}$ and $\Pi'_{h,i}=\Pi'_h(r)=const$, $\hat\Pi_{h,i}=\hat\Pi(r)=const$ in $(r_{i},r_{i+1})$, $i=1,2,\ldots,N$. 



The corresponding penalized functional reads
\begin{align}
\mathcal I_p(\Pi_h,\hat\Pi_h) 
&=\frac{1}{p}\sum_{i=1}^n\Big\{\big\{[
r_{i} + \Pi'_{h,i} + \hat\Pi_{h,i}]^2
+\tilde\ell^{-2}[\Pi^2_h(r_{i}) + (r_{i}\hat\Pi_{h,i}-\Pi_h(r_{i}))^2]\big\}^p+\nonumber\\
&\quad\;\big\{ [r_{i+1} + \Pi'_{h,i} + \hat\Pi_{h,i}]^2
+\tilde\ell^{-2}[\Pi^2_h(r_{i+1}) + (r_{i+1}\hat\Pi_{h,i}-\Pi_h(r_{i+1}))^2]\big\}^p\Big\}\,.
\label{I_Pi_p}
\end{align}

\subsection{Numerical determination of lower and upper bounds of $t^*$}
\label{lower_upper_num_torsion}

Let us recall the problems \eqref{lower_bound9}, \eqref{lower_bound10} and \eqref{I_Pi3}, that is,
\begin{equation}
\tilde t^*=\inf_{\substack{q\in \tilde W\\ \int_0^1 \tilde r^2q\,d\tilde r=1}}\;\int_0^1 \sqrt{q^2+\tilde\ell^2[(q')^2 + (q/\tilde r)^2]}\, \tilde r\,d\tilde r,\quad \tilde W = \{ q \in H^1(0,1)\ |\; q(0) = 0\},
\label{tilde_t*}
\end{equation}
\begin{equation}
\tilde t^*=\frac{1}{\sqrt{\min\limits_{(\Pi,\hat\Pi)\in \hat X_\Pi} \max\limits_{r\in[0,1]}\left\{\left(
		r + \Pi'(r) + \hat\Pi(r))\right)^2
		+\tilde\ell^{-2}\left[\Pi^2(r) + \left(r\hat\Pi(r)-\Pi(r)\right)^2\right]\right\}}}
\label{lower_bound12},
\end{equation}
where $\hat X_\Pi =\big\{(\Pi,\hat\Pi)\in W^{1,\infty}((0,1))\times L^{\infty}((0,1))\ |\;\; \Pi(1)=0\big\}$.
To find lower and upper bounds of $t^*$, we use the penalization methods from Section \ref{sec_penalty}. 

\begin{figure}[!h]
	\centering	
	\includegraphics[width = 0.48\textwidth]{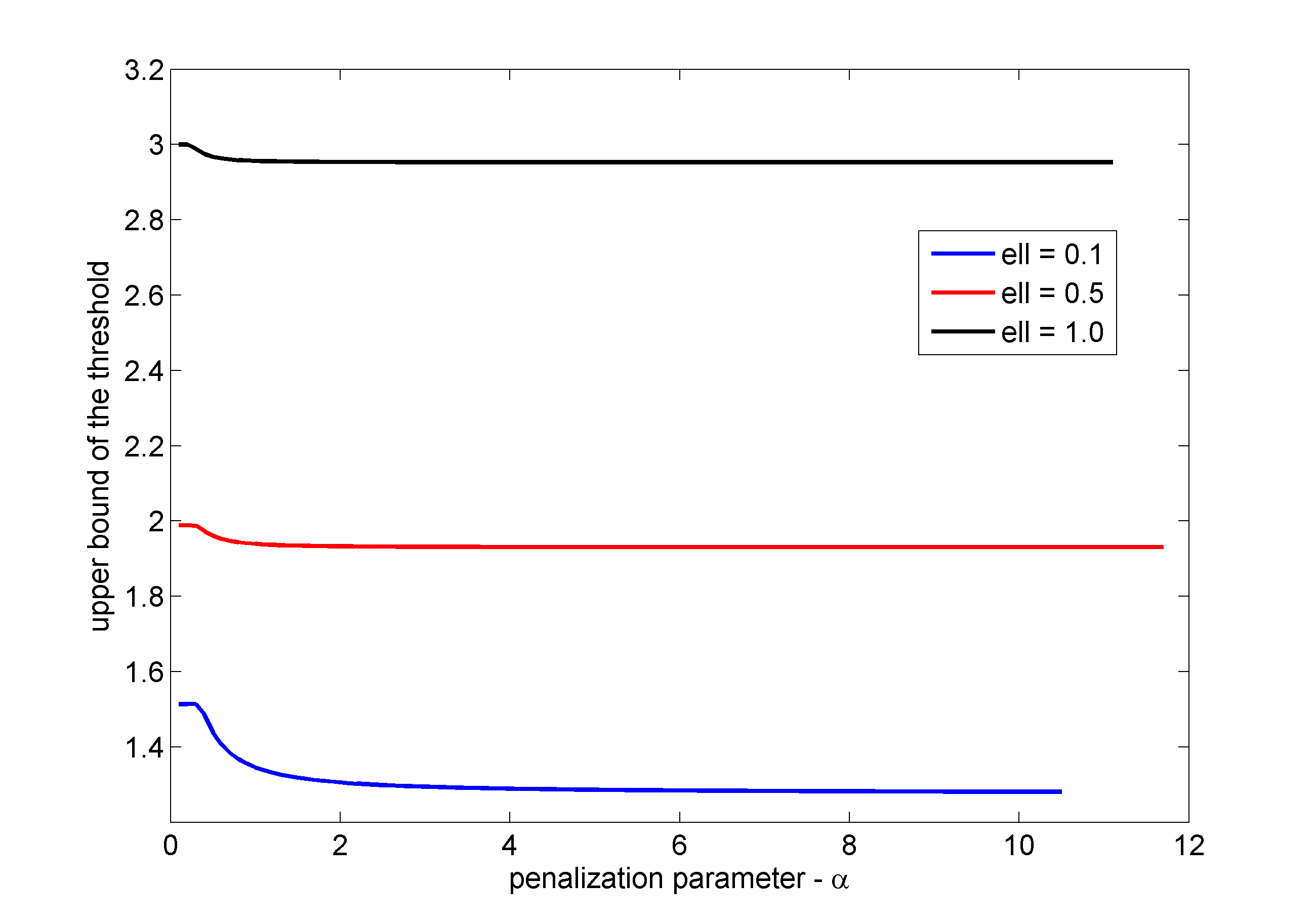}
	\includegraphics[width = 0.48\textwidth]{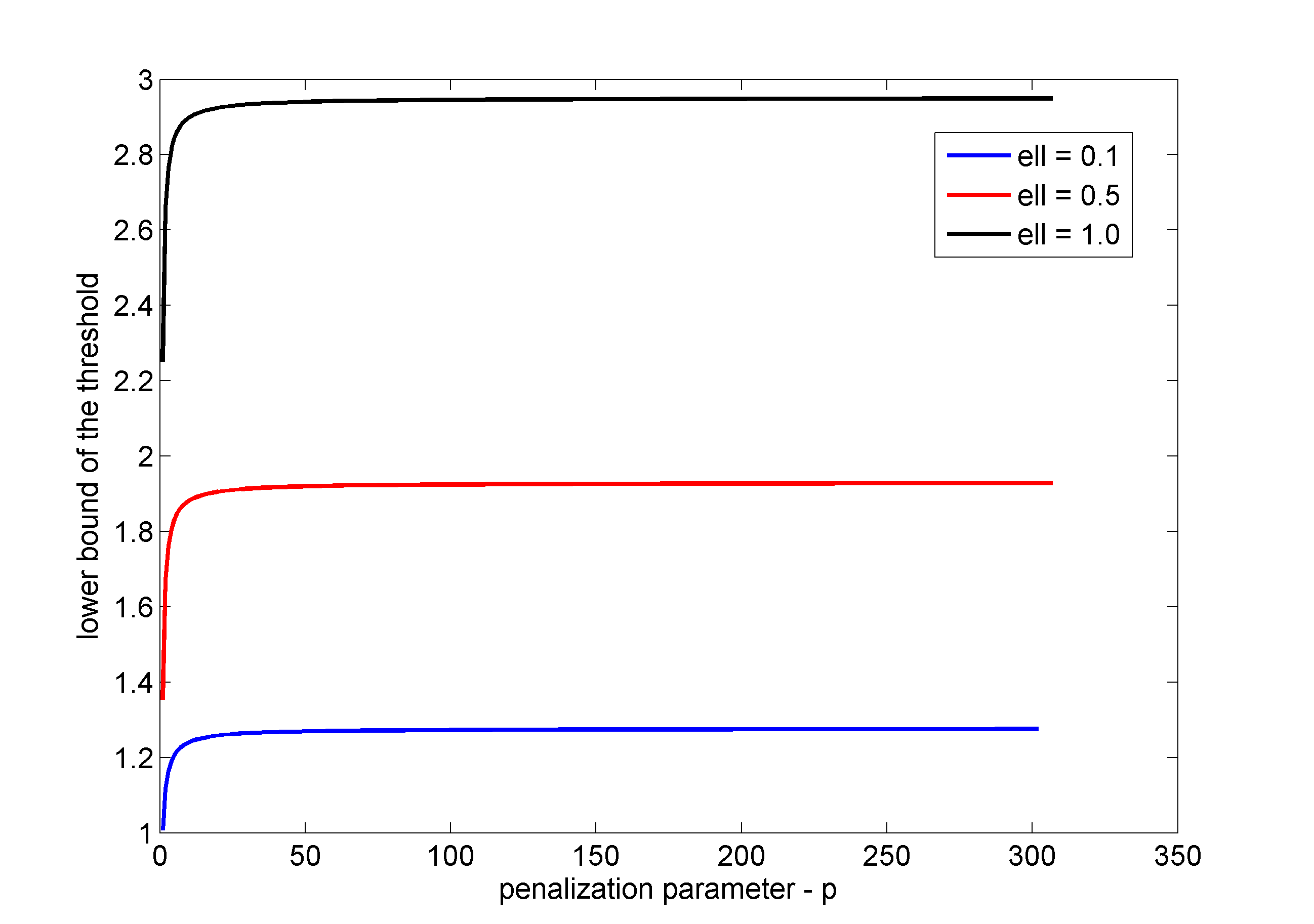}\\
	(a)\hspace{36ex} (b)
\\	\caption{The upper (a) and lower (b) bounds of $\tilde t^*_h$ depending on $p$ for $\tilde\ell=0.1$ (blue), $\tilde\ell=0.5$ (red) and $\tilde\ell=1.0$ (black).}
	\label{fig_torsion3}
\end{figure}

\begin{figure}[!h]
	\centering	
	\includegraphics[width = 0.48\textwidth]{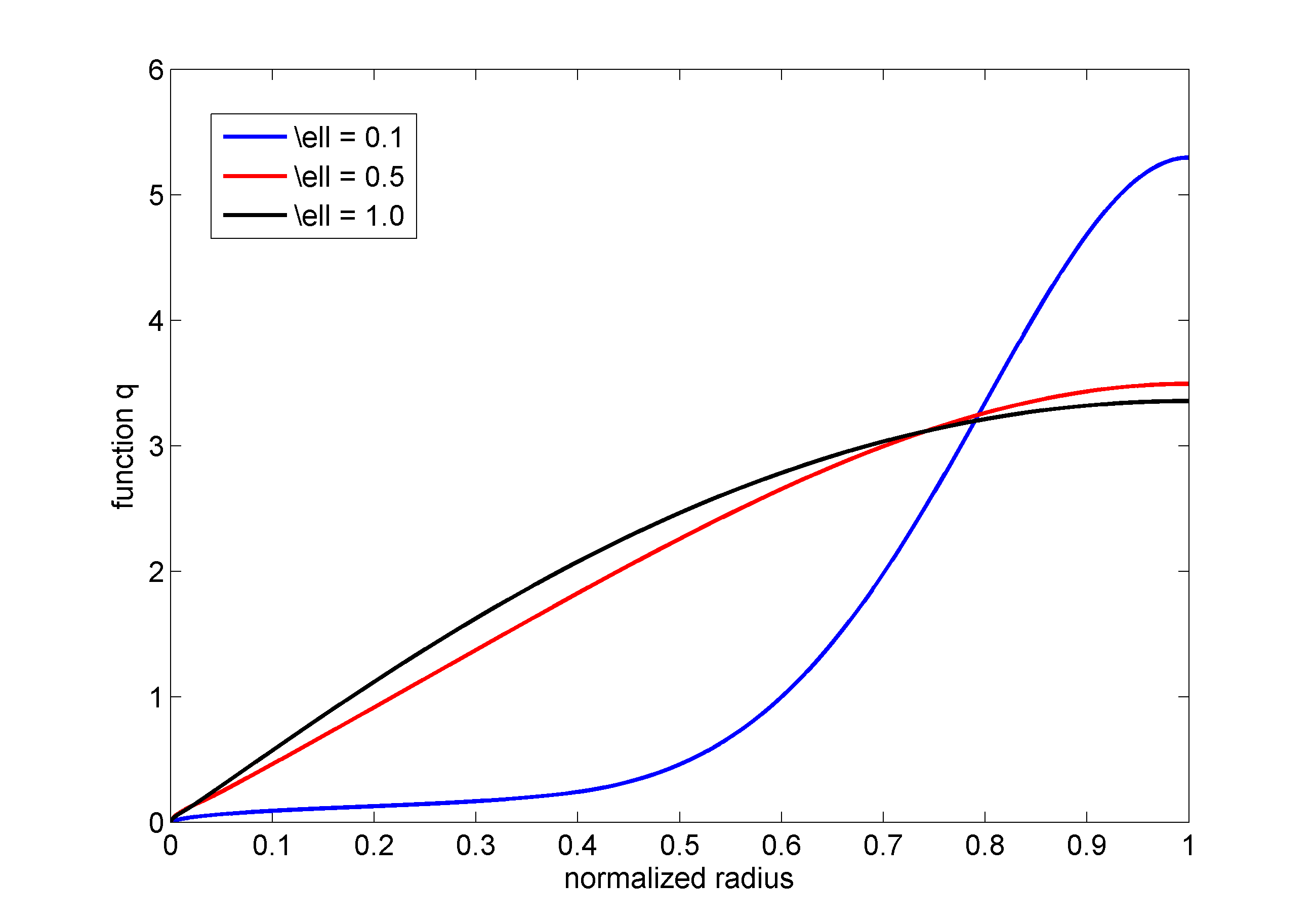}
	\includegraphics[width = 0.48\textwidth]{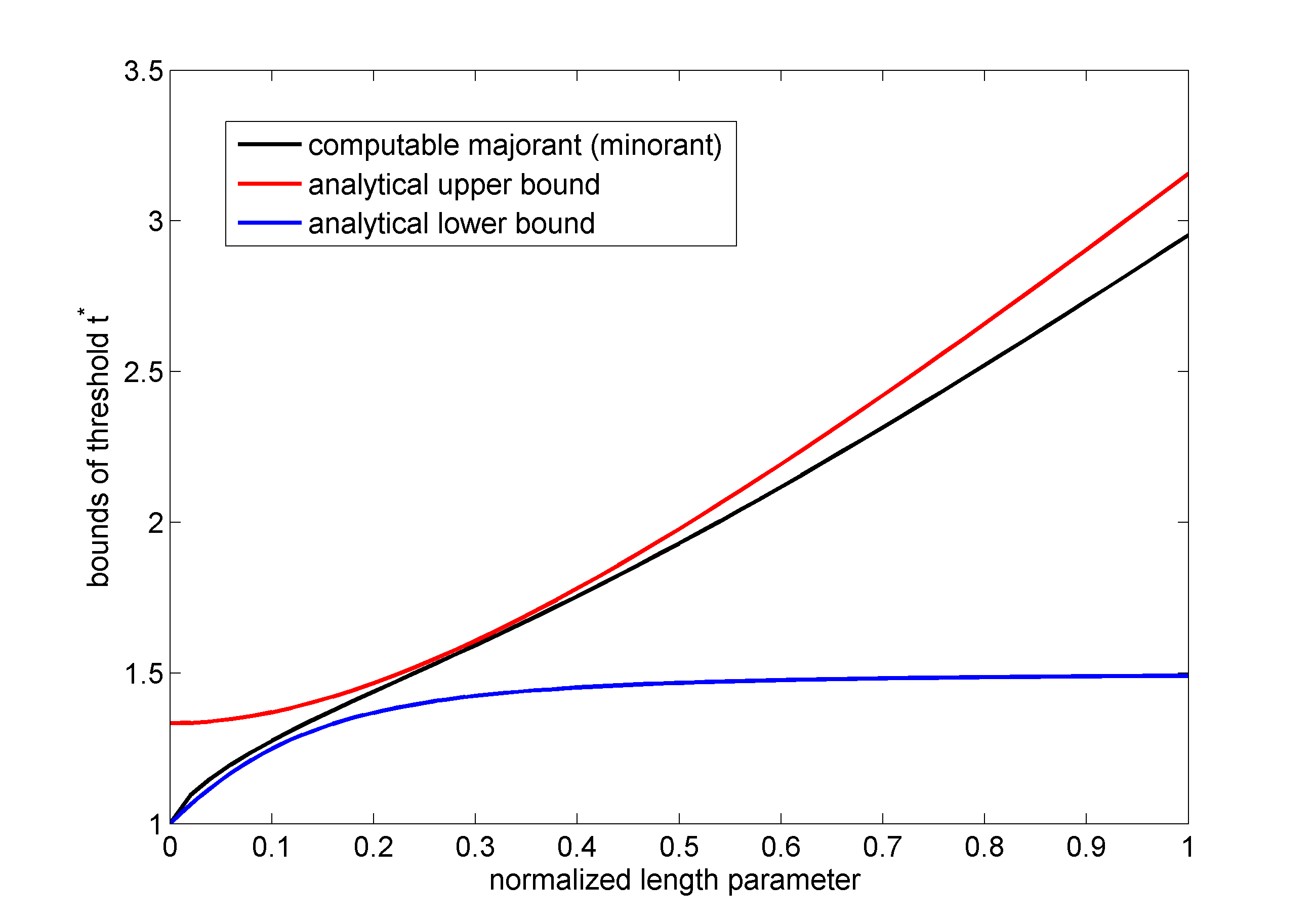}
	\\
	(a)\hspace{36ex} (b)
	\\
	\caption{(a) The optimizing functions $q_\alpha$ for various length scales; (b) comparison of numerical and analytical bounds of $\tilde t^*$}
	\label{fig_torsion}
\end{figure}

\begin{figure}[!h]
	\centering	
	\includegraphics[width = 0.48\textwidth]{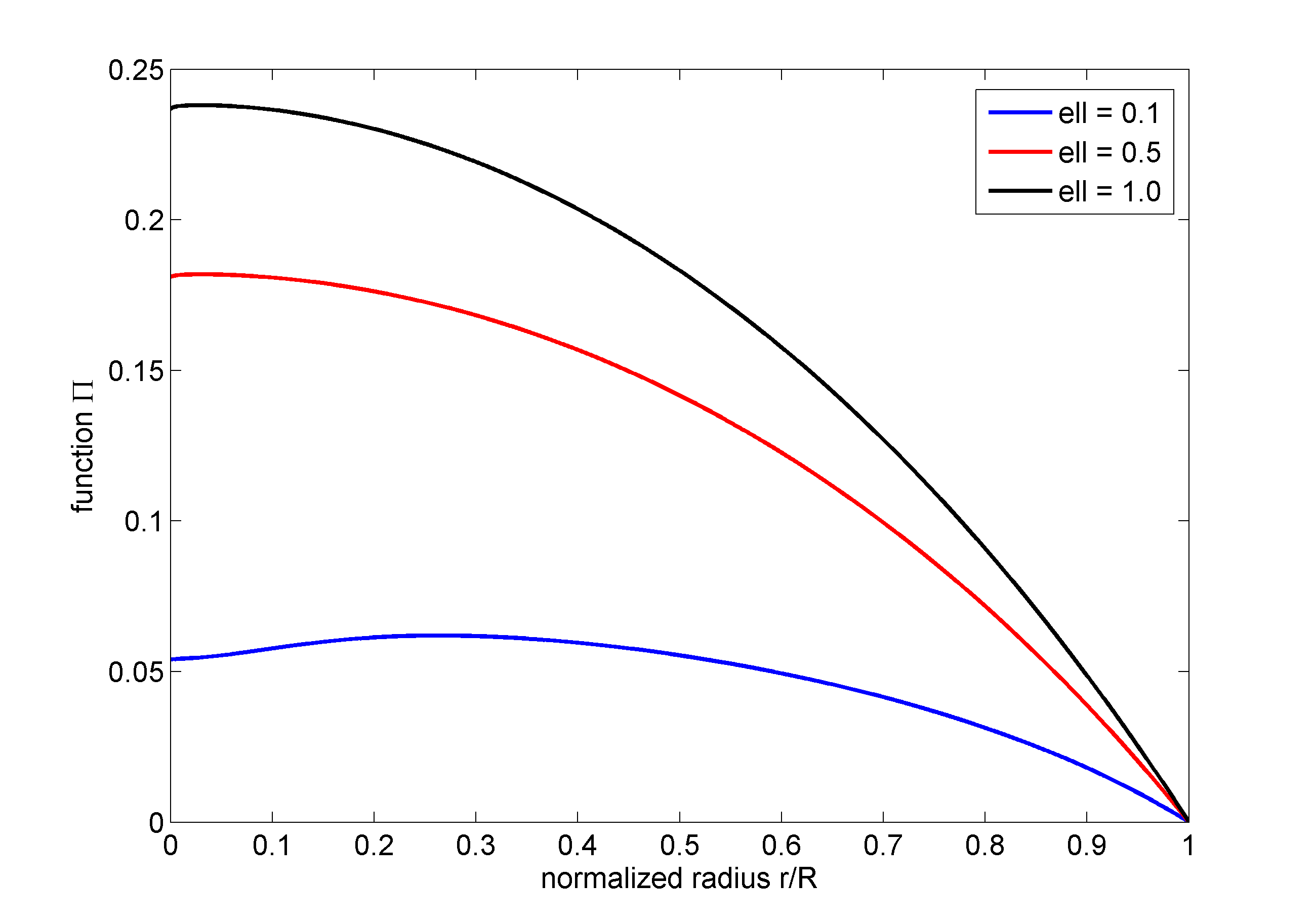}
	\includegraphics[width = 0.48\textwidth]{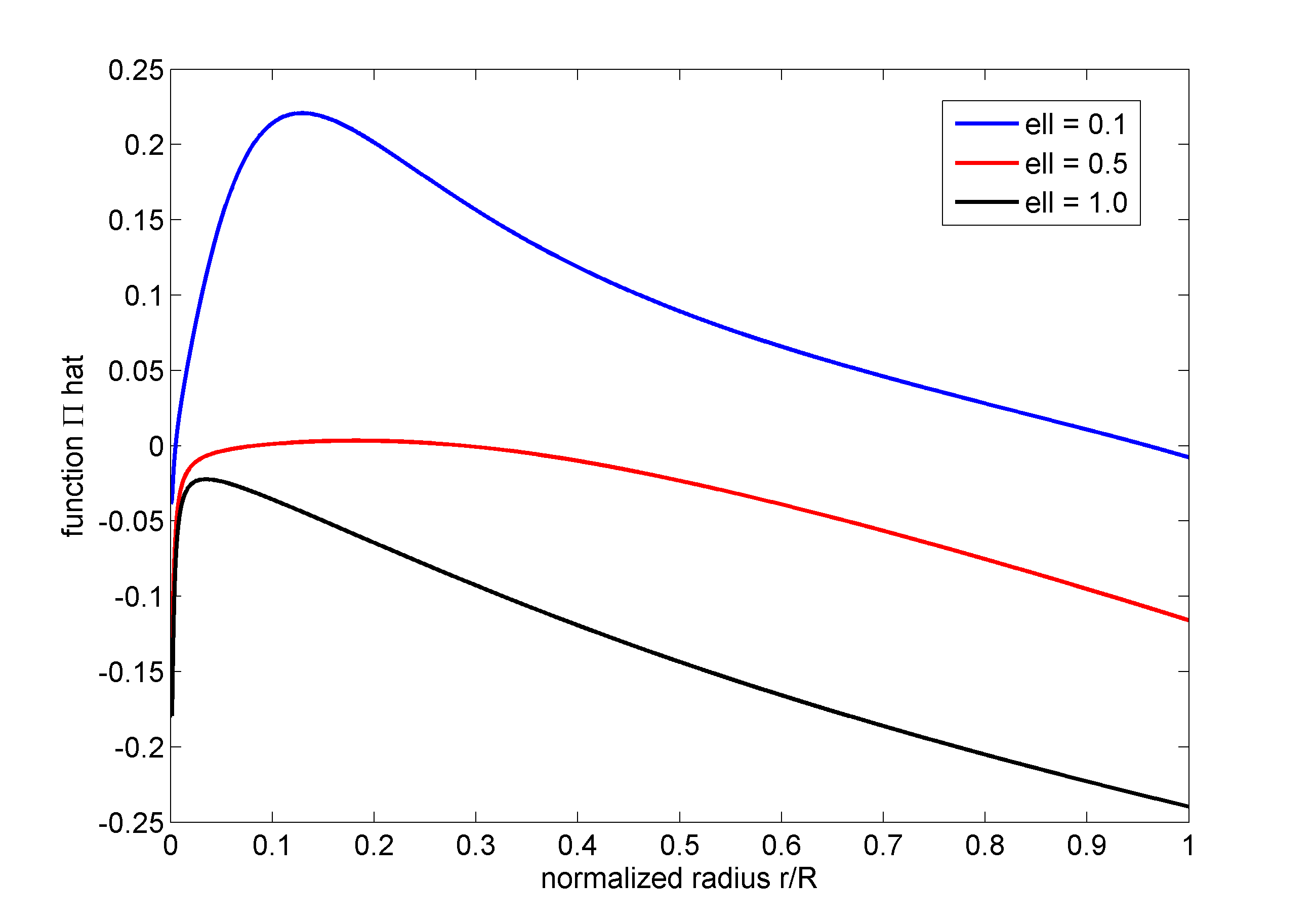}
	\\
	(a) \hspace{36ex} (b) 
	\\
	\caption{The optimizing functions $\Pi$ (a) and $\hat{\Pi}$ (b) for $\tilde\ell=0.1$ (blue), $\tilde\ell=0.5$ (red) and $\tilde\ell=1.0$ (black).}
	\label{fig_torsion2}
\end{figure}

For the upper bound problem \eqref{tilde_t*}, we use piecewise quadratic (P2) elements with 2-point Gauss quadrature and $1000$ elements. By continuation over the penalty parameter $\alpha$ we find that $\alpha\approx10$ is sufficiently large for any $\tilde \ell\in (0,1]$, see Figure \ref{fig_torsion3} (a). 

For the lower bound problem \eqref{lower_bound12} (see also \eqref{lower_bound11}), we use 1000 P1-P0 elements as introduced in Section \ref{subsec_discrete_torsion}. From Figure \ref{fig_torsion3} (b), we see that $p\approx 300$ is a sufficiently large value of the penalization parameter.

Numerical solutions of problems \eqref{tilde_t*} and \eqref{lower_bound12} are depicted in Figure \ref{fig_torsion} (left) and Figure \ref{fig_torsion2} for three different values of $\tilde \ell$. Computable majorants of $\tilde t^*$ following from Figure \ref{fig_torsion3} (a) are 1.2807, 1.9305, and 2.9529 for $\tilde\ell=0.1,\ 0.5,\ 1.0$, respectively. Computable minorants following from Figure \ref{fig_torsion3} (b) are 1.2755, 1.9279, and 2.9493, respectively. We see that the bounds are sufficiently sharp. Figure \ref{fig_torsion} (right) compares computable majorants (minorants) with analytical upper and lower bounds presented in Section \ref{subsec_bounds_anal_torsion}.

\section{Conclusion}

This work has been concerned with the development of a general approach to determining bounds on the load parameter corresponding to incipient plastic behaviour, that is, the elastic threshold, for the dissipative model of strain-gradient plasticity. Such an approach has been motivated by the fact that the yield function, in a generalized associative model, is a function of microstresses and is therefore not a means through which initial yield may be determined while following an elastic path. The approach taken in this work is motivated by the theorems of limit analysis, with bounds being determined through minimization or maximization problems. 

The theory has been applied to two example problems, both of which have received attention in the context of strain-gradient plasticity. For both problems -- the first of a plate in plane stress, and subject to a prescribed displacement, and the second concerning a rod in uniform torsion -- bounds are determined analytically by choosing candidate functions, and numerically by solution of the extremization problems using finite elements. In this latter case the non-differentiable functionals are regularized through the adoption of penalized approximations.

The theory and results presented here shed further light on the behaviour of the dissipative model of strain-gradient plasticity. For example, it is evident that the microstresses in the elastic branch of the loading path are not uniquely defined and their distribution is non-trivial, even if the Cauchy stress is known analytically. 

The techniques of limit analysis have proved to be a powerful tool in the study of this model. It would be useful to consider further applications of this approach, for example to problems of single-crystal plasticity and ensembles of crystals.  



\end{document}

%% file: defs.tex
\newcounter{tony}
\newcommand{\bb}{\mbox{\boldmath{$b$}}}

\newcommand{\bq}{\mbox{\boldmath{$q$}}}

\newcommand{\bu}{\mbox{\boldmath{$u$}}}

\newcommand{\bx}{\mbox{\boldmath{$x$}}}

\newcommand{\bI}{\mbox{\boldmath{$I$}}}

\newcommand{\bT}{\mbox{\boldmath{$T$}}}

\newcommand{\BBR}{\mbox{$\mathbb{R}$}}           
           
\newcommand{\BBC}{\mbox{$\mathbb{C}$}}

\newcommand{\bzero}{\mbox{$\bf 0$}}

\newcommand{\bepsilon}{\mbox{\boldmath{$\varepsilon$}}}

\newcommand{\bpi}{\mbox{\boldmath{$\pi$}}}

\newcommand{\bsigma}{\mbox{\boldmath{$\sigma$}}}

\newcommand{\btau}{\mbox{\boldmath{$\tau$}}}

\newcommand{\bPi}{\mbox{\boldmath{$\Pi$}}}

\newcommand{\beq}{\begin{equation}}

\newcommand{\beqna}{\begin{eqnarray}}
\newcommand{\eeqna}{\end{eqnarray}}